\documentclass[twocolumn,showkeys,showpacs,prb]{revtex4}
\usepackage{amsfonts, amsmath, amssymb, latexsym}
\usepackage[dvips]{graphicx}

\newtheorem{1}{Theorem}
\newtheorem{2}[1]{Theorem}
\newtheorem{3}[1]{Theorem}

\newtheorem{6}{Corollary}

\begin{document}

\title{Asymptotic convergence of the partial averaging technique}

\author{Cristian Predescu}
\author{J. D. Doll}

\affiliation{
Department of Chemistry, Brown University, Providence, Rhode Island 02912
}
\author{David L. Freeman}
\affiliation{Department of Chemistry, University of Rhode Island, Kingston, Rhode Island 02881}
\date{\today}
\begin{abstract}
We study the asymptotic convergence of the partial averaging method, 
a technique used in conjunction with the random series implementation 
of the Feynman-Ka\c{c} formula. We prove asymptotic bounds valid for 
most series representations in the case when the potential has first 
order Sobolev derivatives. If the potential has also second order 
Sobolev derivatives, we prove a sharper theorem which gives  the 
\emph{exact} asymptotic behavior of the density matrices. The results 
are then specialized for the Wiener-Fourier series representation. It 
is found that the asymptotic behavior is 
$o(1/n^2)$ if the potential has first order Sobolev 
derivatives. If the potential has second order Sobolev derivatives, 
the convergence is shown to be ${O}(1/n^3)$ and we give the 
exact expressions for the convergence constants of the density 
matrices.
\end{abstract}
\maketitle

\section{Introduction}
\newcommand{\ud}{\mathrm{d}}
Path integral methods\cite{Fey65} are perhaps the most
important techniques utilized to account for  quantum contributions in
equilibrium statistical simulations for many-body systems. They are based
on the Feynman-Ka\c{c} formula,\cite{Kac50,Sim79} which represents the
density matrix and hence the partition function of a canonical quantum
system as an infinite dimensional stochastic integral of a Brownian
bridge functional.\cite{Pre02}  This stochastic integral is then
approximated by a sequence of finite dimensional integrals, which in turn
are evaluated by Monte Carlo techniques.  The two most important families
of discretization techniques, while not unrelated, can be distinguished
by the way  the paths are parameterized. The discrete path integral (DPI)
methods\cite{Cep95,Ber86,Mie01} utilize the Trotter product
rule\cite{Tro59} and appropriate short-time approximations to construct a
more accurate finite-dimensional approximation of the quantum-mechanical
density matrix. The random series implementations\cite{Dol90,Pre02}
utilize the Ito-Nisio theorem\cite{Kwa92} to construct the Brownian
bridge in the Feynman-Ka\c{c} formula. Independent of which method is
used, a major concern in the practical applications is the asymptotic
rate of convergence as measured against the number of variables  used to
parameterize the paths $n$ as well as the computational time necessary
to evaluate the various quantities involved.

The asymptotic behavior of the discrete methods is well
studied\cite{Suz85, Rae83, Mie01, Pre02c} and is known to be ${O}(1/n^2)$
if the expressions of the short-time approximations are functionals of
the potential only. In contrast, the relative merits of the random series
techniques are poorly understood. There are two basic implementations of
the random series techniques that are important for practical
applications: the partial averaging\cite{Dol85, Pre02} and the reweighted
methods.\cite{Pre02} In this paper, we shall study the asymptotic rates
of convergence of the partial averaging method for different classes of
basis entering the Ito-Nisio theorem as well as for different classes of
potentials. We obtain surprisingly powerful convergence theorems and,
most of the time, we  provide not only the convergence order of different
methods, but also the convergence constants. In the remainder of the
introduction, we try to convince the reader that such an analytical study
is important both theoretically and practically.

The partial averaging method has been rarely utilized in
practical applications because it requires  the Gaussian transform of the
potential for its implementation.  For real life potentials, this is a
difficult but not impossible task. However, it was generally considered
that the improvement the technique brings in does not warrant the effort of
computing the Gaussian transform of the potential and  the so-called
gradient corrected partial averaging method was used instead. It has been
shown that this latter method has general ${O}(1/n^2)$ asymptotic
behavior for sufficiently smooth potentials  and it has been argued that
there is not much reason to suspect a better convergence rate for the
full partial averaging method.\cite{Ele99} However, more accurate
numerical evidence recently presented in Ref.~(\onlinecite{Pre02})
suggests that the full partial averaging method does have in fact a
better behavior: if the technique is used in conjunction with the
Fourier path integral approach (FPI)\cite{Dol84,Dol85} and if the
potential is  smooth enough, the asymptotic convergence  is
${O}(1/n^3)$.

Consistent with these observations, in this paper we show
that the asymptotic rate of convergence for the  FPI partial averaging
approximation (PA-FPI) is indeed ${O}(1/n^3)$ whenever the
potential has second order derivatives in a sense that will be made clear
in the text.  While we shall prove quite general convergence results
valid for almost all series representations, the PA-FPI approach, which
is based on the Wiener-Fourier basis, will receive a special treatment.
This is motivated by the fact that the Wiener-Fourier series
representation is the optimal representation as to the minimization of
the number of variables used to parameterize the paths.\cite{Pre02}

The importance of the partial averaging method resides also
in the fact that it acts as a prototypical strategy for improving the
asymptotic rate of convergence of the random series path integral
methods. As such, the reweighted random series technique\cite{Pre02}
achieves superior asymptotic convergence  by simulating the  partial
averaging approach. For this reason, there is a close relation between
the asymptotic rates of convergence of the two methods. While the
asymptotic behavior of the reweighted method will be the object of a
separate paper, we mention that such a study is greatly simplified once
the  convergence rates of the partial averaging method are known.

The practical importance of the convergence theorems that
will be established in the present paper consists in the fact that they
give \emph{a priori} knowledge about the asymptotic order of convergence
based only on readily verifiable properties of the potential. These
theorems also provide useful estimates of the absolute error in the
evaluation of the density matrices and, by integration, of the partition
functions. Moreover, if the potential has second order derivatives, the
estimates for the density matrix are asymptotically exact and they can be
used to derive the convergence constants for many other properties, as
for instance the convergence constants for the H-method and T-method
energy estimators.\cite{Ele99, Pre02} However, such applications require
careful analytical work that is  beyond the scope of the present paper.

The content of the remainder of this paper is organized
as follows. In the next section, we give a short review of the partial
averaging method with the purpose of establishing the notation. We also
provide a description of the main classes of potentials for which the
asymptotic convergence will be studied. In Section~III, we prove the main
convergence theorems, which are valid for most series representations. We
prove some asymptotic bounds for the density matrix in the case when the
potential has first order Sobolev derivatives and then we give the exact
asymptotic behavior for the potentials having second order Sobolev
derivatives. In Section~IV we specialize the results of Section~III for
the case of the Wiener-Fourier series representation. We prove that if
the potential has first order Sobolev derivatives, then the convergence
is  $o(1/n^2)$. Then, we show that if
the potential has second order Sobolev derivatives, then the convergence
is ${O}(1/n^3)$ and we also provide the exact expression for the
corresponding convergence constant. We illustrate the derived asymptotic
convergence constants for the simple case of the harmonic oscillator and,
in Section~V, we summarize and discuss our results.

\section{The partial averaging strategy}

\subsection{Description of the method}

	In this section, we shall give a short review of
the partial averaging method  with the sole purpose of establishing the
notation. For a more complete discussion, the reader should consult
Ref.~(\onlinecite{Pre02}) and the cited bibliography. The starting point
is the Feynman-Ka\c{c} formula
\begin{equation}
%EQUATION 1
\label{eq:1}
\frac{\rho(x,x';\beta)}{\rho_{fp}(x,x';\beta)}=\mathbb{E}\exp\left\{-\beta\int_{0}^{1}\! 
\!  V\Big[x_r(u)+\sigma B_u^0 \Big]\ud u\right\},
\end{equation}
where $\rho(x,x';\beta)$ is the density matrix for a monodimensional 
canonical system characterized by the inverse temperature 
$\beta=1/(k_B T)$ and made up of identical particles of mass $m_0$ 
moving in the potential $V(x)$.
The stochastic element that appears in Eq.~(\ref{eq:1}), $\{B_u^0,\, u\geq
0\}$, is a so-called standard Brownian bridge defined as follows: if 
$\{B_u,\, u\geq
0\}$ is a standard Brownian motion starting at zero, then the Brownian
bridge is the stochastic process~$\{B_u |\,B_1=0,\, 0 \leq u \leq 1\}$
i.e., a Brownian motion conditioned on~$B_1=0$. In this
paper, we shall reserve the symbol~$\mathbb{E}$ to denote the expected
value (average value) of  a certain random variable against the
underlying probability measure of the Brownian bridge~$B_u^0$. To
complete the description of Eq~(\ref{eq:1}), we set $x_r(u)=x+(x'-x)u$
(called the reference path), $\sigma= (\hbar^2\beta  /m_0)^{1/2}$, and
let $\rho_{fp}(x,x';\beta)$ denote the density matrix for a similar free
particle.

The generalization of the Eq.~(\ref{eq:1}) to a $d$-dimensional
system is straightforward. The symbol $B_u^0$ now denotes a
$d$-dimensional standard Brownian bridge, which is a vector $(B_{u,1}^0,
B_{u,2}^0,\ldots, B_{u, d}^0)$ with the components being independent
standard Brownian bridges. The symbol $\sigma$ stands for the vector
$(\sigma_1, \sigma_2, \ldots, \sigma_d)$ with components defined by
$\sigma_i^2=\hbar^2\beta/m_{0,i}$, while the product $\sigma B_u^0$ is
interpreted as the $d$-dimensional vector of components $\sigma_i B_{u,i}^0$.
Finally, $x$ and $x'$ are points in the configuration space
$\mathbb{R}^d$ while $x_r(u)=x+(x'-x)u$ is a line in $\mathbb{R}^d$
connecting the points $x$ and $x'$. In this paper, we shall conduct the
proofs for monodimensional systems and only state the
easily obtainable $d$-dimensional results. 

	The most general series representation of the Brownian bridge is given
by the Ito-Nisio theorem,\cite{Kwa92} the statement of which is
reproduced below.
Assume given $\{\lambda_k(\tau)\}_{k \geq 1}$
a system of functions on the interval $[0,1]$, which, together
with the constant function
  $\lambda_0(\tau)=1$, makes up an orthonormal basis in $L^2[0,1]$. Let 
\[ \Lambda_k(t)=\int_0^t \lambda_k(u)\ud u.\]
If $\Omega$
is the space of infinite sequences $\bar{a}\equiv(a_1,a_2,\ldots)$ and
\begin{equation}
\label{eq:2}
P[\bar{a}]=\prod_{k=1}^{\infty}\mu(a_k)
\end{equation}
  is the probability measure on $\Omega$ such that the coordinate maps 
$\bar{a}\rightarrow a_k$ are independent identically distributed 
(i.i.d.) variables with distribution probability
\begin{equation}
\label{eq:3}
\mu(a_k\in A)= \frac{1}{\sqrt{2\pi}}\int_A e^{-z^2/2}\,\ud z,
\end{equation}
then
\begin{equation}
\label{eq:4}
B_u^0(\bar{a}) = \sum_{k=1}^{\infty}a_k\Lambda_{k}(u),\; 0\leq u\leq1
\end{equation}
is equal in distribution to a standard Brownian bridge. Moreover,
the convergence of the above series is almost surely uniform on the
interval $[0,1]$.

Using the Ito-Nisio representation of the Brownian bridge,
the Feynman-Ka\c{c} formula (\ref{eq:1}) takes the form
\begin{eqnarray}
\label{eq:5}
  \frac{\rho(x, x' ;\beta)}{\rho_{fp}(x, x' 
;\beta)}&=&\int_{\Omega}\ud P[\bar{a}]\nonumber  \exp\bigg\{-\beta 
\int_{0}^{1}\! \!  V\Big[x_r(u) \\& +& \sigma \sum_{k=1}^{\infty}a_k 
\Lambda_k(u) \Big]\ud u\bigg\}.
\end{eqnarray}
The independence of the coordinates $a_k$, which physically
amounts to choosing those representations in which the kinetic energy
operator is diagonal, is the key to the use of the partial averaging
method. Denoting by $\mathbb{E}_{n}$ the average over the coefficients
beyond the rank~$n$, the partial averaging formula reads
\begin{eqnarray}
\label{eq:6}
  &&\frac{\rho_n^{{PA}}(x, x' ;\beta)}{\rho_{fp}(x, x' 
;\beta)}=\int_{\mathbb{R}}\ud \mu(a_1)\ldots \int_{\mathbb{R}}\ud 
\mu(a_n)\nonumber \\ &&\times \exp\bigg\{-\beta \; 
\mathbb{E}_n\int_{0}^{1}\! \!
V\Big[x_r(u)+\sigma \sum_{k=1}^{\infty}a_k
\Lambda_k(u) \Big]\ud u\bigg\}. \qquad
\end{eqnarray}

To make more sense of the above formula, it is convenient to use the
notation introduced in Ref.~(\onlinecite{Pre02})
\[S_u^n(\bar{a})=\sum_{k=1}^n a_k \Lambda_k(u) \quad \text{and} \quad 
B_u^n(\bar{a})=\sum_{k=n+1}^\infty a_k \Lambda_k(u), \]
which denote the  $n$th order partial sum and and $n$th order tail
series, respectively.
By construction, the random variables $S_u^n(\bar{a})$ and 
$B^n_u(\bar{a})$ are independent and
\[B_u^0(\bar{a})=\sum_{k=1}^\infty a_k 
\Lambda_k(u)=S_u^n(\bar{a})+B_u^n(\bar{a}).\]
Moreover, the variables $B_u^n$ and $B_\tau^n$ have a joint
Gaussian distribution of covariances
\begin{eqnarray}\nonumber
\label{eq:7}&&
\mathbb{E}_n(B_u^n)^2=\sum_{k=n+1}^\infty \Lambda_k(u)^2, \quad 
\mathbb{E}_n(B_\tau^n)^2=\sum_{k=n+1}^\infty 
\Lambda_k(\tau)^2,\nonumber \\&&\text{and} \quad 
\gamma_n(u,\tau)=\mathbb{E}_n(B_u^nB_\tau^n)=\sum_{k=n+1}^\infty 
\Lambda_k(u)\Lambda_k(\tau). \qquad \end{eqnarray}
Equivalently, by using the fact that $S_u^n(\bar{a})$ and 
$B^n_u(\bar{a})$ are independent and their sum is  $B^0_u(\bar{a})$, 
one may evaluate the above series to be
\begin{eqnarray*}
\mathbb{E}_n(B_u^n)^2=\mathbb{E}(B_u^0)^2-\sum_{k=1}^n 
\Lambda_k(u)^2=u(1-u)-\sum_{k=1}^n \Lambda_k(u)^2,\\ 
\mathbb{E}_n(B_\tau^n)^2=\mathbb{E}(B_\tau^0)^2-\sum_{k=1}^n 
\Lambda_k(\tau)^2=\tau(1-\tau)-\sum_{k=1}^n \Lambda_k(\tau)^2,
\end{eqnarray*}
and
\begin{eqnarray*}&&
\gamma_n(u,\tau)=\mathbb{E}(B_u^0B_\tau^0)-\sum_{k=1}^n 
\Lambda_k(u)\Lambda_k(\tau)\\&&=\min(u,\tau)-u\tau-\sum_{k=1}^n 
\Lambda_k(u)\Lambda_k(\tau) .\end{eqnarray*}
The computation of $\mathbb{E}(B_u^0)^2$, $\mathbb{E}(B_\tau^0)^2$,
and $\mathbb{E}(B_u^0B_\tau^0)$ is 
straightforward and  is performed in Appendix~A.

Going back to Eq.~(\ref{eq:6}),  one inverts the order of
integration in the exponent and computes
\begin{eqnarray}
\label{eq:8}&&
\mathbb{E}_n\!\int_0^1\! \ud u V[x_r(u)+\sigma 
B_u^0(\bar{a})]\nonumber \\&& = \int_0^1\! \ud u \mathbb{E}_n 
V[x_r(u)+\sigma S_u^n(\bar{a})+\sigma B_u^n(\bar{a})] 
\\&&=\int_0^1\!  \overline{V}_{u,n}[x_r(u)+\sigma S_u^n(\bar{a})]\ud 
u, \nonumber
\end{eqnarray}
where
\begin{equation}
\label{eq:9}
\overline{V}_{u,n}(y)=\int_{\mathbb{R}}\frac{1}{\sqrt{2\pi\Gamma_{n}^2(u)}} 
\exp\left[-\frac{z^2}{2\Gamma_{n}^2(u)}\right]V(y+z) \ud z,
\end{equation}
with~$\Gamma_n^2(u)$ defined by
\begin{equation}
\label{eq:10}
\Gamma_{n}^2(u)=\sigma^2 \mathbb{E}_n(B_u^n)^2=
\sigma^2\left[u(1-u)-
\sum_{k=1}^{n}\Lambda_k(u)^2\right].
\end{equation}
In deducing Eq.~(\ref{eq:8}), one uses the fact that the
variable $B_u^n$ has a Gaussian distribution of variance
$\mathbb{E}_n(B_u^n)^2$. To summarize, we define the $n$th order partial
averaging approximation to the diagonal density matrix by the formula
\begin{eqnarray}
\label{eq:11}&&
\frac{\rho^{PA}_n(x, x' ;\beta)}{\rho_{fp}(x, x' 
;\beta)}=\int_{\mathbb{R}}\ud \mu(a_1)\ldots \int_{\mathbb{R}}\ud 
\mu(a_n)\nonumber  \\&& \times \exp\bigg\{-\beta \; \int_{0}^{1}\! \!
\overline{V}_{u,n}\Big[x_r(u)+\sigma
\sum_{k=1}^{n}a_k \Lambda_k(u) \Big]\ud u\bigg\}.\qquad
\end{eqnarray}

It has been shown\cite{Pre02} that the rate of convergence of the
partial averaging method is  controlled to a first approximation by the
rate of decay of $\Gamma^2_n(u)$ to zero. More precisely, the  value
$\int_{0}^1 \Gamma^2_n(u) \ud u$ may be taken as a measure of the
efficiency of different basis sets $\{\lambda_k(\tau)\}_{k \geq 1}$. It
was found that $\int_{0}^1 \Gamma^2_n(u) \ud u$ attains its unique minimum
\[
\sigma^2\left(\frac{1}{6}-\sum_{k=1}^n \frac{1}{\pi^2 k^2}\right)
\] on the cosine-Fourier basis $ \{\lambda_k(\tau) =\sqrt{2} \cos(k 
\pi \tau);\; k \geq 1\}$. The corresponding series representation for 
the Brownian bridge is  the Wiener-Fourier construction
\[
B_u^0(\bar{a})\stackrel{d}{=} 
\sqrt{\frac{2}{\pi^2}}\sum_{k=1}^{\infty}a_k\frac{\sin({k\pi 
u})}{k},\; 0\leq u\leq1
\]
and was used by Doll and Freeman\cite{Dol84} in their
definition of the Fourier Path Integral method. Though their work was
based on arguments other than those presented here, we consider that the
name of Wiener-Fourier Path Integral (WFPI) method is more appropriate
because the random series
\[
a_0u+\sqrt{\frac{2}{\pi^2}}\sum_{k=1}^{\infty}a_k\frac{\sin({k\pi 
u})}{k},\; 0\leq u\leq1
\] was historically the first explicit construction of a standard 
Brownian motion  and was due to Wiener.\cite{Wie23}

\subsection{Mathematical considerations}
In this subsection, we discuss the classes of
basis $\{\lambda_k(u)\}_{k\geq 1}$ as well as the classes of smooth
enough potentials for which  we study the asymptotic convergence.
Anticipating later results from this paper, the rate of convergence of
the partial averaging sequence of approximations of the density matrix
will be shown to depend solely on the properties of the function
$\gamma_n(u,\tau)$ defined by Eq.~(\ref{eq:7}). Regarded as an integral kernel,  the
function $\gamma_n(u,\tau)$ defines a positive definite quadratic form on
$L^2([0,1])$, that is
\[
\int_0^1 \int_0^1 \gamma_n(u,\tau)f(u)f(\tau)\ud u \ud \tau \geq 
0,\quad \forall f \in L^2([0,1]).
\]
It is but a simple exercise to verify that in
general $\gamma_n(u,\tau)^k$ defines such a positive definite quadratic
form for all integers $k \geq 1$. In particular, we are interested in the
convergence properties of $\gamma_n(u,\tau)^2$.

In this paper, we shall restrict our discussion to those
basis $\{\lambda_k(u)\}_{k\geq 1}$ for which there is a distribution
$g_2(u,\tau)$ such that (i)
\begin{equation}
\label{eq:12}
\lim_{n \to \infty} \frac{\gamma_n(u,\tau)^2}{\int_0^1 \int_0^1 
\gamma_n(u,\tau)^2 \ud u \ud \tau} = g_2(u,\tau)
\end{equation}
in the sense of distributions and (ii) $g_2(u,\tau)$ defines an 
integral kernel on $L^2([0,1])$ that is strictly positive definite. 
By convergence in the sense of distributions, we understand
\begin{eqnarray*}
\lim_{n \to \infty} \frac{\int_0^1 \int_0^1 \gamma_n(u,\tau)^2 
h(u,\tau)\ud u \ud \tau }{\int_0^1 \int_0^1 \gamma_n(u,\tau)^2 \ud u 
\ud \tau} \\= \int_0^1 \int_0^1 g_2(u,\tau) h(u,\tau) \ud u \ud \tau
\end{eqnarray*}
for all continuous functions $h(u,\tau)$, while the strictly positive 
definiteness of $g_2(u,\tau)$ is the statement
\[
\int_0^1 \int_0^1 g_2(u,\tau)f(u)f(\tau)\ud u \ud \tau = 0 \iff f=0 
\; \text{a.s.}
\]

We mention that it is not clear if in fact these two
conditions are satisfied for all basis $\{\lambda_k(u)\}_{k\geq 1}$.
However, the reader may always verify them for the problem at hand. We
prove in Appendix~B that for the Wiener-Fourier basis
\begin{equation}
\label{eq:13}
\lim_{n \to \infty} \frac{\gamma_n(u,\tau)^2}{\int_0^1 \int_0^1 
\gamma_n(u,\tau)^2 \ud u \ud \tau} = \delta(u-\tau)
\end{equation}
in the sense of distributions. Moreover, $\delta(u-\tau)$ is indeed 
strictly positive definite because
\[
\int_0^1 \int_0^1 \delta(u-\tau)f(u)f(\tau)\ud u \ud \tau =\int_0^1 
f(u)^2\ud u
\]
is zero if and only if $f=0$ a.s.

A special category of basis $\{\lambda_k(u)\}_{k\geq 1}$ are
those for which the primitives $\Lambda_k(u)$ do not change sign. An
example is furnished by the Haar wavelet basis, which generates the so
called L\'evy-Ciesielski representation of the Brownian bridge.\cite{Pre02c, McK69} For such
basis, the functions $\gamma_n(u,\tau)$ are positive and we shall further
assume that there is a distribution $g_1(u,\tau)$ such that (i)
\begin{equation}
\label{eq:14}
\lim_{n \to \infty} \frac{\gamma_n(u,\tau)}{\int_0^1 \int_0^1 
\gamma_n(u,\tau)\ud u \ud \tau} = g_1(u,\tau)
\end{equation}
in the sense of distributions and (ii) $g_1(u,\tau)$ defines
an integral kernel on $L^2([0,1])$ that is strictly positive definite. We
analyze this category of basis separately because there are  more general
convergence results available for them.

In the second part of this subsection, we shall discuss the
class of potentials for which the convergence is analyzed.  For sure, the analysis must be restricted to those potentials for which the Feynman-Ka\c{c} formula holds and for which the method does in fact converge.  The Feynman-Ka\c{c} formula~(\ref{eq:1}) is known to hold for a quite large class of potentials which is called the Kato class.  On the other hand, it has been proven\cite{Pre02b} that the partial
averaging method is convergent  for all series
representations and for all Kato-class potentials that have finite
Gaussian transform. We provide below the mathematical definition of this class. The readers who  find this definition meaningless might get a better insight about the nature of the potentials in this class by studying the examples we later provide.    To arrive at the definition of the Kato
class,\cite{Aiz82} we let
\[
g(y)=\left\{\begin{array}{cc} |y| &  d=1, \\ \ln({\|y\|^{-1}})& d=2, 
\\ \|y\|^{2-d}& d\geq 3, \end{array}\right.
\]
and define the Kato class $K_d$ as the set of all measurable 
functions $f: \mathbb{R}^d \to \mathbb{R}$ such that
\begin{equation}
\label{eq:15}
\lim_{\alpha \downarrow 0} \sup_{x\in \mathbb{R}^d} \int_{\|x-y\|\leq 
\alpha}|f(y)g(x-y)|\ud y =0.
\end{equation}
We also say that $f$ is in $K_{d}^{\text{loc}}$ if $1_{D}f \in K_{d}$ 
for all bounded domains $D \subset \mathbb{R}^d$. We say that  $V(x)$ 
is a Kato-class potential if its negative part $V_{-}=\max\{0, -V\}$ 
is in $K_d$ while its positive part $V_{+}=\max\{0,V\}$ is in 
$K_d^{\text{loc}}$.

Roughly speaking, Eq.~(\ref{eq:15}) imposes a restriction on the 
singularities of the potentials. As far as the chemical physicist is concerned, the standard prototype as to the nature of the singularities is the  general $n$-body potential
\begin{equation}
\label{eq:15a}
V(\bar{r}_1,\ldots, \bar{r}_n)= \sum_{i=1}^n\frac{a_i}{\|\bar{r}_i-\bar{R}_i\|^q}+ \sum_{i,j}^n \frac{b_{i,j}}{\|\bar{r}_i-\bar{r}_j\|^q}, 
\end{equation}
where the individual particles are assumed to move in the usual tridimensional space.
Using the observation that both $K_d$ and $K_d^{\text{loc}}$ classes are linear spaces and also using Proposition~4.3 of Ref.~(\onlinecite{Aiz82}), the reader may prove that the potentials of the form given by Eq.~(\ref{eq:15a}) are in the class $K_d$ for all $q<2$.  Such potentials include for instance the coulombic potentials as they appear in electronic structure calculations and for which $q=1$. Remark that the singularities in Eq.~(\ref{eq:15a}) can be oriented upward or downward. It does not make any difference as to the validity of the Feynman-Ka\c{c} formula as well as the convergence of the partial averaging method.  

 Another important example of Kato-class potentials are the functions $V(x)$ that are continuous and bounded from below.  It is trivial to verify that if $c_1 \geq 0$ and $c_2\geq 0$ are some nonnegative numbers and $V_1(x)$ and  $V_2(x)$ are Kato-class potentials, then $c_1V_1(x)+c_2V_2(x)$ is also a Kato-class potential. Using this observation, one may argue that any ab initio potential obtained at the level of the Born-Oppenheimer approximation must be a Kato-class potential. This is so because it can usually be written as a sum between a $K_d$ potential of the type given by Eq.~(\ref{eq:15a}) (the internuclear repulsion term) and a continuous and bounded from below potential (this term arises from the electron-nuclear  and electron-electron interactions). More generally, even for strongly ionic systems, one notices that the Born-Oppenheimer potential can have at most coulombic negative singularities   and therefore it is of Kato-class.  
 
Many of the properties of the Kato-class potentials were enumerated by Aizenman and Simon.\cite{Aiz82} As argued by them, the Kato-class is the natural and in many aspects maximal class of potentials for which the Feynman-Ka\c{c} formula holds.  It is known for instance that any Kato-class potential is locally integrable. That is,  
\[
\int_D |V(x)| \ud x =\int_{\mathbb{R}^d} 1_D |V(x)| \ud x< \infty
\]
for all bounded domains $D\in \mathbb{R}^d$. Because of this restriction, the Leonard-Jones potential as well as some other empirical potentials are not included in the Kato-class due to the $r^{-12}$ singularity.  Even if the Feynman-Ka\c{c} formula may be well defined for such potentials with positive non-integrable singularities,  it is not clear if the computed answer is the Green's function of the corresponding Bloch equation. However, as far as the chemical physicist is concerned such issues are hardly of any relevance because the Kato class comprises all potentials of physical interest. The Leonard-Jones potential can be brought into the Kato-class by truncation or other approximations.

The condition that the potentials have finite Gaussian transform 
can be formulated as follows. For all vectors 
$\alpha=(\alpha_1,\ldots,\alpha_d)\in \mathbb{R}_+^d$ of strictly 
positive components, we consider the Gaussian measure
\begin{equation}
\label{eq:16}
\ud \mu_\alpha(z)= \left(\prod_{i=1}^d \frac{1}{{2\pi 
\alpha_{i}^2}}\right)^{1/2}\exp\left(-\sum_{i=1}^d 
\frac{z_i^2}{2\alpha_{i}^2}\right) \ud z_1 \cdots \ud z_d.
\end{equation}
Then, a potential is said to have finite Gaussian transform if
\begin{eqnarray}
\label{eq:17}
\overline{|V|}_{\alpha}(x)=\int_{\mathbb{R}^d}  \left|V(x+z)\right| 
\ud \mu_\alpha(z) < \infty,
\end{eqnarray}
for all $x\in \mathbb{R}^d$ and $\alpha \in \mathbb{R}_+^d$. Let us notice that from the practical point of view, this additional condition is hardly a restriction. The potentials used in actual simulations are usually a sum between the Born-Oppenheimer interatomic potential and a \emph{constraining} potential. The first one always has a finite Gaussian transform. The second one is added in order to ensure that the partition function is finite
\[
Z(\beta)<\infty, \quad \forall \, \beta>0
\]
and is intended to simulate, for example, the effect of the container in which a reaction takes place. For this purpose one usually uses a continuous and bounded from below potential, which is, of course, a Kato-class potential. If, for instance, the potential increases to infinity slower than any exponential $\exp({c \|x\|^2})$ with $c>0$, then 
it has a finite Gaussian transform. But this is hardly a restriction since the typical constraining potentials are polynomials. 

Under these conditions, it was shown\cite{Pre02b}
that the partial averaging sequence of approximations is convergent to
the correct Feynman-Ka\c{c} result (which by itself is properly defined) at least as far as the convergence of
the density matrix and the partition function is concerned. However, we
expect that the rate of convergence depends on the smoothness of the
potential. As mentioned in the introduction,  we are interested in
establishing the class of potentials for which the fastest convergence is
achieved. While the maximal class depends upon the specific series
representation, we shall see that there is a sufficiently large class of
potentials for which  all series representations achieve their fastest
asymptotic convergence.

 Theorem~3.b) of
Ref.~(\onlinecite{Pre02b}) says that in order to verify that a potential
has a finite Gaussian transform, it is enough to verify that
\begin{eqnarray}
\label{eq:18}
\overline{|V|}_{\alpha}(x_0)=\int_{\mathbb{R}^d} 
\left|V(x_0+z)\right| \ud \mu_\alpha(z) < \infty,
\end{eqnarray}
for all $\alpha \in \mathbb{R}_+^d$ and an arbitrarily given  $x_0\in 
\mathbb{R}^d$. If we choose $x_0=0$, then the set of potentials 
having finite Gaussian transform is the set $\cap_\alpha L^1_\alpha 
(\mathbb{R}^d)$, where  $L^1_\alpha (\mathbb{R}^d)$ is the space of 
functions $f$ for which the weighted norm
\begin{eqnarray}
\label{eq:19}
\|f\|_{1,\alpha}=\int_{\mathbb{R}^d}  \left|f(z)\right| \ud \mu_\alpha(z)
\end{eqnarray}
is finite. By Theorem~3.a) of Ref.~(\onlinecite{Pre02b}), $L^1_\alpha 
(\mathbb{R}^d)\subset L^1_{\text{loc}}(\mathbb{R}^d)$ and conversely, 
one may argue that a function has finite Gaussian transform if it is 
locally integrable and its modulus increases slower than any Gaussian 
at infinity.

However, in this paper we shall assume that the potential $V(x)$ lies 
in the set  $\cap_\alpha L^2_\alpha (\mathbb{R}^d)$, where 
$L^2_\alpha (\mathbb{R}^d)$ is the space of functions $f$ for which 
the weighted norm
\begin{eqnarray}
\label{eq:20}
\|f\|_{2,\alpha}=\left[\int_{\mathbb{R}^d}  f(z)^2 \ud \mu_\alpha(z) 
\right]^{1/2}
\end{eqnarray}
is finite. Since the measure $\ud \mu_\alpha(z)$ is a probability 
measure, one may apply the Cauchy-Schwartz inequality and see that
$
\|f\|_{1,\alpha}\leq \|f\|_{2,\alpha},
$
so that $L^2_\alpha (\mathbb{R}^d)\subset L^1_\alpha (\mathbb{R}^d)$ 
and $\cap_\alpha L^2_\alpha (\mathbb{R}^d)\subset \cap_\alpha 
L^1_\alpha (\mathbb{R}^d)$. Therefore, the class of potentials 
discussed in the present paragraph have finite Gaussian transform. We 
remind the reader that $L^2_\alpha (\mathbb{R}^d)\subset 
L^2_{\text{loc}} (\mathbb{R}^d)$ and that the spaces $L^2_\alpha 
(\mathbb{R}^d)$ are Hilbert spaces.

Still, the class of potentials introduced in
the previous paragraph is not smooth enough for the purpose of studying
the asymptotic convergence. Later in the paper, we shall see  that
natural classes of smooth enough potentials are intersections of weighted
Sobolev spaces $\cap_\alpha W^{m,2}_\alpha(\mathbb{R}^d)$ for $m=1$ or
$m=2$. Since the functions $f(x) \in L^2_\alpha (\mathbb{R}^d)$ are
locally integrable,  they have  partial distributional derivatives of any
order. The weighted Sobolev space $W^{m,2}_\alpha(\mathbb{R}^d)$ is the
space of all functions  $f\in L^2_\alpha (\mathbb{R}^d)$ whose partial
distributional  derivatives up to the order $m$ are also  $L^2_\alpha
(\mathbb{R}^d)$ functions. In an equivalent definition, one may argue
that $f\in \cap_\alpha W^{m,2}_\alpha(\mathbb{R}^d)$ if and only if
\[
f(x)\exp\left(-\sum_{i=1}^d \frac{x_i^2}{2\alpha_{i}^2}\right) \in 
W^{m,2}(\mathbb{R}^d),
\]
for all $\alpha \in \mathbb{R}^d_+$, where $W^{m,2}(\mathbb{R}^d)$ is 
the usual Sobolev space on $\mathbb{R}^d$.

There are two main categories of Kato-class
potentials which we study in this paper: those that also lie in
$\cap_\alpha W^{1,2}_\alpha(\mathbb{R}^d)$ and, more restrictively, those
that lie in $\cap_\alpha W^{2,2}_\alpha(\mathbb{R}^d)$. For the first
class of potentials, the weighted Sobolev norms
\[
\int_{\mathbb{R}^d} \left\{V(x)^2+\sum_{i=1}^d [\partial_i 
V(x)]^2\right\} \ud \mu_\alpha(x)
\]
are finite, while for the second class of potentials, the norms
\[
\int_{\mathbb{R}^d} \left\{V(x)^2+\sum_{i=1}^d [\partial_i 
V(x)]^2+\sum_{i,j}^d[\partial_{i,j} V(x)]^2\right\} \ud \mu_\alpha(x)
\]
are finite.  These requirements imposes further restrictions on the 
singularities of the Kato-class potentials. We leave it for the 
reader to show by explicit computation that the potentials of the form given by Eq.~(\ref{eq:15a})  lie in the $\cap_\alpha W^{1,2}_\alpha(\mathbb{R}^{3n})$ 
space if and only if $q < 1/2$ and  in the $\cap_\alpha 
W^{2,2}_\alpha(\mathbb{R}^{3n})$ space if and only if $q < - 1/2$. For another example, a general $3n$-body potential of the form given by the Eq.~(\ref{eq:15a}) but where the inter-particle potential is replaced with the Morse potential  lies also in the $\cap_\alpha 
W^{2,2}_\alpha(\mathbb{R}^{3n})$ space.

\section{Asymptotic convergence of the partial averaging method}
In this section, we shall establish several  general convergence theorems
valid for all series representations and for smooth potentials. While for most of the paper we  adopt a rigorous mathematical style, we shall not
provide a formal proof of the key conjecture expressed by Eqs.~(\ref{eq:25}) and (\ref{eq:25a}).
Instead, we shall try to present arguments to support this
plausible assertion within reasonable doubt and then explore its
implications.

 We start by defining
\[
U_n(x,x',\beta; \bar{a})=\int_0^1\overline{V}_{u,n}[x_r(u)+\sigma 
\sum_{k=1}^n a_k \Lambda_k(u)] \ud u
\]
and
\[
U_\infty(x,x',\beta; 
\bar{a})=\int_0^1V[x_r(u)+\sigma\sum_{k=1}^\infty a_k \Lambda_k(u) ] 
\ud u.
\]
The variables $x$, $x'$, and $\beta$ are interpreted here
as parameters and, just as a reminder, we separate them 
from the ``true'' variable $\bar{a}$ by a semicolon. We notice that
$U_n(x,x',\beta; \bar{a})$ as a function of $\bar{a}$ depends only
on the first $n$ coefficients $a_1, \ldots, a_n$. Moreover, in terms of
these functions, the relation (\ref{eq:8}) reads \begin{equation}
\label{eq:21}
U_n(x,x',\beta;\bar{a})=\mathbb{E}_n 
\left[U_{\infty}(x,x',\beta;\bar{a})\right]
\end{equation}
and in fact,
\begin{equation}
\label{eq:22} \lim_{n \to \infty} U_n(x,x',\beta;\bar{a})= 
U_{\infty}(x,x',\beta;\bar{a}),
\end{equation}
as shown by Theorem~1 of Ref.~(\onlinecite{Pre02b}).
To continue with the introduction of the notations, we define
\[X_n(x,x',\beta;\bar{a})=\rho_{fp}(x,x';\beta)\exp[-\beta \, 
U_n(x,x',\beta;\bar{a})]\]
and
\[X_\infty(x,x',\beta;\bar{a})=\rho_{fp}(x,x';\beta)\exp[-\beta\, 
U_\infty(x,x',\beta;\bar{a})].\]
Then, we have
\begin{equation}
  \label{eq:23} \rho(x,x';\beta)=\mathbb{E}\,[X_\infty(x,x',\beta;\bar{a})]
\end{equation}
and
\begin{equation}
\label{eq:24} 
\rho_{n}^{\text{PA}}(x,x';\beta)=\mathbb{E}\,[X_n(x,x',\beta;\bar{a})],
\end{equation}
respectively.

A little algebra shows that
\begin{eqnarray*}
\mathbb{E}_n 
X_\infty(x,x',\beta;\bar{a})-X_n(x,x',\beta;\bar{a})=X_n(x,x',\beta;\bar{a})\\ 
\times \,\mathbb{E}_n\left\{e^{-\beta 
[U_{\infty}(x,x',\beta;\bar{a})-U_n(x,x',\beta;\bar{a})]}-1\right\}.
\end{eqnarray*}
However, for large $n$, we can expand the exponential in a Taylor 
series and retain the first non-vanishing positive term in the 
series, which is also the one that controls the asymptotic 
convergence. We have:
\begin{eqnarray}
\label{eq:25} \nonumber
\mathbb{E}_n 
X_\infty(x,x',\beta;\bar{a})-X_n(x,x',\beta;\bar{a})\approx 
X_n(x,x',\beta;\bar{a})\\ \times \frac{\beta^2}{2}\mathbb{E}_n 
\left[U_{\infty}(x,x',\beta;\bar{a})-U_n(x,x',\beta;\bar{a})\right]^2.\quad
\end{eqnarray}
 From now on, if $A_n> 0$ and $B_n > 0$, we interpret $A_n\approx B_n$ to mean
\[
\lim_{n \to \infty} {A_n}/{B_n} =1,
\]
while $A_n \lessapprox B_n$ is interpreted to mean
\[
\limsup_{n \to \infty} A_n/B_n \leq 1.
\]
The error in the equation (\ref{eq:25}) is of the order
\[
\frac{\beta^3}{3!}\mathbb{E}_n 
\left|U_{\infty}(x,x',\beta;\bar{a})-U_n(x,x',\beta;\bar{a})\right|^3,
\]
which decays at a faster rate than the local variance
\begin{equation*}
\mathbb{E}_n 
\left[U_{\infty}(x,x',\beta;\bar{a})-U_n(x,x',\beta;\bar{a})\right]^2.
\end{equation*}
Therefore, the use of the symbol $\approx$ is justified any time the 
local variance is not zero. However, this property is true for almost 
all $\bar{a}$ whenever the potential is not constant. In 
Eq.~(\ref{eq:25}), the  term of order one in the Taylor expansion 
cancels because of the identity (\ref{eq:21}), so the asymptotic 
behavior is dictated by the local variance of the function 
$U_\infty(x,x',\beta;\bar{a})$. The Eq.~(\ref{eq:25}) is expected to 
be true for all potentials $V(x)\in \cap_\alpha 
L^2_\alpha(\mathbb{R})$ which are not constant and, from now on, we 
shall assume that the potential satisfies these criteria.
Taking the total expectation $\mathbb{E}$ in Eq.~(\ref{eq:25}) and remembering that $X_n(x,x',\beta;\bar{a})$ does not depend upon the coefficients beyond the rank $n$, we obtain
\begin{eqnarray}
\label{eq:25a} \nonumber
\rho(x,x';\beta)-\rho_n^{\text{PA}}(x,x';\beta)\approx\frac{\beta^2}{2}\mathbb{E}\,\big\{ 
X_n(x,x',\beta;\bar{a})\\ \times 
\left[U_{\infty}(x,x',\beta;\bar{a})-U_n(x,x',\beta;\bar{a})\right]^2 \big\}.\quad
\end{eqnarray}

\subsection{Potentials having  first order Sobolev derivatives}

If $V(x)\in \cap_\alpha W^{1,2}_\alpha(\mathbb{R})$, 
Eq.~(\ref{eq:C7}) and Eq.~(\ref{eq:C9}) of  Appendix~C show that
\begin{eqnarray}
\label{eq:26}\nonumber
\mathbb{E}_n T'_n(x,x',\beta;\bar{a})&\leq&
\mathbb{E}_n 
\left[U_{\infty}(x,x',\beta;\bar{a})-U_n(x,x',\beta;\bar{a})\right]^2 
\\& \leq& \mathbb{E}_n T_n(x,x',\beta;\bar{a}),
\end{eqnarray}
where $T_n(x,x',\beta;\bar{a})$ and $T_n(x,x',\beta;\bar{a})$ are \emph{positive} 
functions defined by the equations
\begin{eqnarray}
\label{eq:27}
\nonumber
T'_n(x,x',\beta;\bar{a})=\sigma^2 \int_0^1 \! \ud u \! \int_0^1\! \ud 
\tau \,  \gamma_n(u,\tau)  \overline{V}'_{u,n}[x_r(u)\\ +\sigma 
S_{u}^n(\bar{a})]  \overline{V}'_{\tau,n}[x_r(\tau) +\sigma 
S^n_\tau(\bar{a})]
\end{eqnarray}
and
\begin{eqnarray}
\label{eq:28}
\nonumber
T_n(x,x',\beta;\bar{a})=\sigma^2 \int_0^1 \! \ud u \! \int_0^1\! \ud 
\tau \,  \gamma_n(u,\tau)  V'[x_r(u)\\ +\sigma B_{u}^0(\bar{a})] 
V'[x_r(\tau) +\sigma B^0_\tau(\bar{a})],
\end{eqnarray}
respectively.
Here, $V'(x)$ denotes the first order derivative of $V(x)$, while 
$\overline{V}'_{u,n}(x)$ is the first order derivative of 
$\overline{V}_{u,n}(x)$.

With the help of the functions $T'_n(x,x',\beta;\bar{a})$ and 
$T_n(x,x',\beta;\bar{a})$ and of  the equation (\ref{eq:25}) one may 
write
\begin{eqnarray}
\label{eq:29} &&\nonumber \frac{\beta^2}{2}\mathbb{E}_n\left[ 
X_n(x,x',\beta;\bar{a}) T'_n(x,x',\beta;\bar{a})\right]\\&& 
\lessapprox
\mathbb{E}_n X_\infty(x,x',\beta;\bar{a})-X_n(x,x',\beta;\bar{a}) 
\\&& \lessapprox \frac{\beta^2}{2}\mathbb{E}_n\left[ 
X_n(x,x',\beta;\bar{a}) T_n(x,x',\beta;\bar{a})\right], \nonumber
\end{eqnarray}
where we used the fact that the function $ X_n(x,x',\beta;\bar{a}) $ 
does not depend upon the coefficients beyond the rank $n$ and thus, 
it can be placed inside the $E_n$ sign. Now, taking the total 
expectation $\mathbb{E}$ in the above equation, we obtain
\begin{eqnarray}&&
\label{eq:30} \nonumber \frac{\beta^2}{2}\mathbb{E}\left[ 
X_n(x,x',\beta;\bar{a}) T'_n(x,x',\beta;\bar{a})\right]\\&& 
\lessapprox \rho(x,x';\beta)-\rho_n^{\text{PA}}(x,x';\beta)
  \\&& \lessapprox \frac{\beta^2}{2}\mathbb{E}\left[ 
X_n(x,x',\beta;\bar{a}) T_n(x,x',\beta;\bar{a})\right].\nonumber
\end{eqnarray}
Moreover, due to  the relation (\ref{eq:22}), we have \[\Delta 
X_n(x,x',\beta;\bar{a})= 
X_n(x,x',\beta;\bar{a})-X_\infty(x,x',\beta;\bar{a}) \to 0\] as $n\to 
\infty$ and  we expect that
\[
\lim_{n \to \infty}\frac{ \mathbb{E}\left[ \Delta 
X_n(x,x',\beta;\bar{a})T_n(x,x',\beta;\bar{a}) 
\right]}{\mathbb{E}\left[ 
X_\infty(x,x',\beta;\bar{a})T_n(x,x',\beta;\bar{a}) \right]}=0,
\]
because the function $T_n(x,x',\beta;\bar{a})$ is positive while 
$X_\infty(x,x',\beta;\bar{a}) $ is strictly positive. Therefore, the 
decay of $\Delta X_n(x,x',\beta;\bar{a})$ to zero makes the numerator 
converge to zero at a faster rate than the denominator.
A  relation similar to the one above holds for the function 
$T'_n(x,x',\beta;\bar{a})$ and consequently, we may replace the 
estimate (\ref{eq:30}) with
\begin{eqnarray}&&
\label{eq:31} \nonumber
  \frac{\beta^2}{2}\mathbb{E}\left[ X_\infty(x,x',\beta;\bar{a}) 
T'_n(x,x',\beta;\bar{a})\right]
\\&& \lessapprox\rho(x,x';\beta)-\rho_n^{\text{PA}}(x,x';\beta)
  \\&& \lessapprox \frac{\beta^2}{2}\mathbb{E}\left[ 
X_\infty(x,x',\beta;\bar{a}) T_n(x,x',\beta;\bar{a})\right].\nonumber
\end{eqnarray}

With these preparations, we are able to state the following theorem:
\begin{1}
\label{th:1}
Assume $V(x)$ is a Kato-class potential that lies in $\cap_\alpha 
W^{1,2}_\alpha (\mathbb{R})$. Then,
\begin{eqnarray}
\label{eq:32} \nonumber &&
\rho(x,x';\beta)-\rho_n^{\text{PA}}(x,x';\beta)\\&& \lessapprox 
\frac{\beta^2}{2}  \int_0^1 \! \ud u \! \int_0^1\! \ud \tau 
\gamma_n(u,\tau)K_{x,x'}^\beta(u,\tau),
\end{eqnarray}
where
\begin{eqnarray}
\label{eq:33}\nonumber
K_{x,x'}^\beta(u,\tau)=\sigma^2\mathbb{E}\Big\{ 
X_{\infty}(x,x',\beta;\bar{a})V'[x_r(u)+\sigma B_{u}^0(\bar{a})]\\ 
\quad \times V'[x_r(\tau)+\sigma B_{\tau}^0(\bar{a})]\Big\}.\quad
\end{eqnarray}

If in addition $\gamma_n(u,\tau)\geq 0$, then the following stronger 
result holds
\begin{eqnarray}
\label{eq:34} \nonumber
\rho(x,x';\beta)-\rho_n^{\text{PA}}(x,x';\beta) \approx 
\frac{\beta^2}{2} \left[ \int_0^1 \! \ud u \! \int_0^1\! \ud \tau 
\gamma_n(u,\tau)\right]\\ \times\left[ \int_0^1 \! \ud u \! 
\int_0^1\! \ud \tau  g_1(u,\tau)K_{x,x'}^\beta(u,\tau)\right].\quad
\end{eqnarray}
\end{1}

\emph{Proof.} The first part of the theorem follows directly  from 
Eqs.~(\ref{eq:27}) and (\ref{eq:31}), by mere substitution. In order 
to prove the second part of the theorem, one divides the terms in 
Eq.~(\ref{eq:31}) by
\[\int_0^1 \! \ud u \! \int_0^1\! \ud \tau \gamma_n(u,\tau)\]
and then uses Eq.~(\ref{eq:14}) as well as the expressions 
(\ref{eq:27}) and (\ref{eq:28}) to show that
\begin{widetext}
\begin{eqnarray}
\label{eq:35}\nonumber
\frac{\beta^2}{2}\lim_{n \to \infty} \frac{ \mathbb{E}\left[ 
X_\infty(x,x',\beta;\bar{a}) 
T'_n(x,x',\beta;\bar{a})\right]}{\int_0^1 \! \ud u \! \int_0^1\! \ud 
\tau \gamma_n(u,\tau)} = \lim_{n \to \infty}
\frac{\rho(x,x';\beta)-\rho_n^{\text{PA}}(x,x';\beta)}{\int_0^1 \! 
\ud u \! \int_0^1\! \ud \tau \gamma_n(u,\tau)}
\\= \frac{\beta^2}{2}\lim_{n \to \infty} \frac{ \mathbb{E}\left[ 
X_\infty(x,x',\beta;\bar{a}) T_n(x,x',\beta;\bar{a})\right]}{\int_0^1 
\! \ud u \! \int_0^1\! \ud \tau \gamma_n(u,\tau)} =  \int_0^1 \! \ud 
u \! \int_0^1\! \ud \tau  g_1(u,\tau)K_{x,x'}^\beta(u,\tau).
\end{eqnarray}
\end{widetext}
  In this respect, one should notice that the use of Eq.~(\ref{eq:14}) 
is justified by the fact that $K_{x,x'}^\beta(u,\tau)$ is a 
continuous function in the arguments $(u,\tau)$. By construction, 
$K_{x,x'}^\beta(u,\tau)$ is symmetric under the change of variables 
$u$ and $\tau$, so it is enough to restrict its analysis to the case 
$\tau \leq u$. However, if $\tau \leq u$, we have
\begin{eqnarray}\nonumber
\label{eq:36}
K_{x,x'}^\beta(u,\tau)=\sigma^2\int_{\mathbb{R}}\int_{\mathbb{R}} 
\rho(x,y;\tau\beta)\rho[y,z;(u-\tau)\beta] \\ 
\times\rho[z,x';(1-u)\beta] V'(y)V'(z)\,\ud y\, \ud z.
\end{eqnarray}
The continuity of $K_{x,x'}^\beta(u,\tau)$ then follows from the 
continuity of the density matrix $\rho(x,x';\beta)$ in all arguments 
and from the fact that the integrals appearing in Eq.~(\ref{eq:36}) 
are finite if $V(x)\in \cap_\alpha W^{1,2}_\alpha (\mathbb{R})$.

Equation~(\ref{eq:35})
implies the relation (\ref{eq:34}) provided that the common limit is not
zero or infinity. We have
\[\int_0^1 \! \ud u \! \int_0^1\! \ud \tau 
g_1(u,\tau)K_{x,x'}^\beta(u,\tau)\leq \sup_{u,\tau} 
K_{x,x'}^\beta(u,\tau)<\infty,\] where we used the fact that the 
integral of $ g_1(u,\tau)$ on the set $[0,1]\times [0,1]$ is $1$ as 
well as the fact that $K_{x,x'}^\beta(u,\tau)$ is continuous on the 
same (compact) set, thus bounded. On the other hand, using 
Eq.~(\ref{eq:33}) and the fact that the kernel $g_1(u,\tau)$ is 
strictly positive definite, the equality
\begin{eqnarray*}
\int_0^1 \! \ud u \! \int_0^1\! \ud \tau 
g_1(u,\tau)K_{x,x'}^\beta(u,\tau)= \sigma^2 \mathbb{E}
\bigg\{ X_{\infty}(x,x',\beta;\bar{a})\\ \times \int_0^1 \! \ud u \! 
\int_0^1\! \ud \tau  g_1(u,\tau)V'[x_r(u)+\sigma B_{u}^0(\bar{a})]\\ 
\quad \times V'[x_r(\tau)+\sigma B_{\tau}^0(\bar{a})]\bigg\}=0
\end{eqnarray*}
implies $V'[x_r(u)+\sigma B_{u}^0(\bar{a})]=0$ for almost all 
$\bar{a}$. However, this is not possible because it contradicts the 
fact  that the potential $V(x)$ is not constant. Therefore, the limit 
in Eq.~(\ref{eq:35}) is finite and non-zero and the proof of the 
theorem is concluded. $\quad \Box$

Let us notice that the asymptotic rate of convergence for the partial 
averaging method depends upon the nature of the basis 
$\{\lambda_k(u)\}_{k\geq 1}$ solely through the behavior of the 
kernels $\gamma_n(u,\tau)$. The kernel $K_{x,x'}^\beta(u,\tau)$ is 
independent of the particular basis and has the important property
\begin{equation}
\label{eq:37}
K_{x,x'}^\beta(1-u,1-\tau)=K_{x',x}^\beta(u,\tau).
\end{equation}
The relation (\ref{eq:37}) can be proven as follows (without loss of 
generality, one may assume $\tau \leq u$):
\begin{eqnarray*}
K_{x,x'}^\beta(1-u,1-\tau)=\sigma^2\int_{\mathbb{R}}\int_{\mathbb{R}} 
\rho[x,y;(1-u)\beta] \\ 
\times\rho[y,z;(u-\tau)\beta]\rho(z,x';\tau\beta) 
V^{(1)}(y)V^{(1)}(z)\,\ud y\, \ud 
z\\=\sigma^2\int_{\mathbb{R}}\int_{\mathbb{R}} 
\rho(x',z;\tau\beta)\rho[z,y;(u-\tau)\beta] \rho[y,x;(1-u)\beta] \\ 
\times V^{(1)}(y)V^{(1)}(z)\,\ud y\, \ud z =K_{x',x}^\beta(\tau, 
u)=K_{x',x}^\beta(u,\tau),
\end{eqnarray*}
where we used the symmetry of the various density matrices appearing 
in the above formula.

The relation (\ref{eq:33}) remains true for $d$-dimensional systems 
provided that we set $\sigma_i^2=\hbar^2\beta/(2 m_{0,i})$ and 
redefine the kernel $K_{x,x'}^\beta(u,\tau)$ to be
\begin{eqnarray}\nonumber
\label{eq:38}
K_{x,x'}^\beta(u,\tau)=\int_{\mathbb{R}^d}\int_{\mathbb{R}^d} 
\rho(x,y;\tau\beta)\rho[y,z;(u-\tau)\beta] \\ 
\times\rho[z,x';(1-u)\beta]\,\left\{ \sum_{i=1}^d 
\sigma_i^2\left[\partial_i V(y)\,\partial_i V(z)\right]\right\}\,\ud 
y\, \ud z.
\end{eqnarray}

\subsection{Potentials having  second order Sobolev derivatives}

Theorem~1 provides a useful bound for the asymptotic rate of 
convergence, but gives the exact asymptotic convergence only if 
$\gamma_n(u,\tau)\geq 0$. In general, if these last functions are not 
positive, additional cancellations might occur if the potential is 
smooth enough. In this case, we need to establish lower and upper 
bounds that are sharper than those offered by Eq.~(\ref{eq:26}).

It turns out that the natural class of potentials on which all series 
representations attain their fastest convergence is the class 
$\cap_\alpha W^{2,2}_\alpha (\mathbb{R})$. For potentials $V(x)$ in 
this class,  Eqs.~(\ref{eq:C8}) and (\ref{eq:C10}) of Appendix~C 
provide  the following bounds
\begin{eqnarray}
\label{eq:39}\nonumber
\mathbb{E}_n Y'_n(x,x',\beta;\bar{a})&\lessapprox &
\mathbb{E}_n 
\left[U_{\infty}(x,x',\beta;\bar{a})-U_n(x,x',\beta;\bar{a})\right]^2\\& 
\lessapprox &\mathbb{E}_n Y''_n(x,x',\beta;\bar{a}),
\end{eqnarray}
where $Y'_n(x,x',\beta;\bar{a})$ and $Y''_n(x,x',\beta;\bar{a})$ are 
\emph{positive} functions defined by the equations
\begin{eqnarray}
\label{eq:40}
\nonumber
Y'_n(x,x',\beta;\bar{a})=T'_n(x,x',\beta;\bar{a})+\frac{\sigma^4}{2} 
\int_0^1 \! \ud u \! \int_0^1\! \ud \tau \, \gamma_n(u,\tau)^2  \\ 
\times \overline{V}_{u,n}''[x_r(u)+\sigma 
S_{u}^n(\bar{a})]\overline{V}_{\tau,n}''[x_r(\tau)+\sigma 
S^n_\tau(\bar{a})]\quad
\end{eqnarray}
and
\begin{equation}
\label{eq:40a}
Y''_n(x,x',\beta;\bar{a})=\frac{1}{2}\left[T'_n(x,x',\beta;\bar{a})+T_n(x,x',\beta;\bar{a})\right],
\end{equation}
respectively. Here, $\overline{V}_{t,n}''(x)$ denotes the second 
derivative of the averaged potential $\overline{V}_{t,n}(x)$.  By 
replacing  the functions $Y'_n(x,x',\beta;\bar{a})$ and 
$Y''_n(x,x',\beta;\bar{a})$ in Eq.~(\ref{eq:25}) and taking the total 
expectation $\mathbb{E}$, one  deduces the following sharper version 
of Eq.~(\ref{eq:30})
\begin{eqnarray}&&
\label{eq:40b} \nonumber \frac{\beta^2}{2}\mathbb{E}\left[ 
X_n(x,x',\beta;\bar{a}) Y'_n(x,x',\beta;\bar{a})\right]\\&& 
\lessapprox \rho(x,x';\beta)-\rho_n^{\text{PA}}(x,x';\beta)
  \\&& \lessapprox \frac{\beta^2}{2}\mathbb{E}\left[ 
X_n(x,x',\beta;\bar{a}) Y''_n(x,x',\beta;\bar{a})\right].\nonumber
\end{eqnarray}
In deducing Eq.~(\ref{eq:40b}), one should remember that  the 
functions $X_n(x,x',\beta;\bar{a})$ do not depend upon the 
coefficients $a_k$ with $k\geq n+1$ and, therefore, $\mathbb{E}\,[X_n 
Y'_n]=\mathbb{E}\{X_n\, \mathbb{E}_n[Y'_n]\}$ and $\mathbb{E}\,[X_n 
Y''_n]=\mathbb{E}\{X_n \,\mathbb{E}_n[Y''_n]\}$.

In the next few paragraphs we prove that the signs $\lessapprox$ in 
Eq.~(\ref{eq:40b}) can be replaced by $\approx$ and then we prove an 
analog of Eq.~(\ref{eq:31}).
As shown by Eqs.~(\ref{eq:C11}) and (\ref{eq:C12}) of Appendix C,
\begin{eqnarray}
\label{eq:40e}&&\nonumber
\mathbb{E}_n Y''_n(x,x',\beta;\bar{a})- \mathbb{E}_n 
Y'_n(x,x',\beta;\bar{a}) \leq  \frac{\sigma^4}{2} \int_0^1 \! \ud u 
\! \int_0^1\! \ud \tau \,\\&& \times  \gamma_n(u,\tau)^2
\bigg\{\mathbb{E}_n\big\{{V}''[x_r(u)+\sigma B_{u}^0(\bar{a})] 
{V}''[x_r(\tau)\\&&+\sigma B^0_\tau(\bar{a})]\big\} - 
\overline{V}_{t,n}''[x_r(u)+\sigma 
S_{u}^n(\bar{a})]\overline{V}_{t,n}''[x_r(\tau)+\sigma 
S^n_\tau(\bar{a})]\bigg\}. \nonumber
\end{eqnarray}
Using this relation together with the Eq.~(\ref{eq:12}), one readily 
proves that
\begin{eqnarray*}
&&\nonumber
\lim_{n \to \infty}\bigg\{\frac{\mathbb{E}\left[ 
X_n(x,x',\beta;\bar{a}) Y''_n(x,x',\beta;\bar{a})\right]}{\int_0^1 \! 
\ud u \! \int_0^1\! \ud \tau \gamma_n(u,\tau)^2}\\&&-\frac{\mathbb{E}\left[ 
X_n(x,x',\beta;\bar{a}) Y'_n(x,x',\beta;\bar{a})\right]}{\int_0^1 \! 
\ud u \! \int_0^1\! \ud \tau \gamma_n(u,\tau)^2}\bigg\}=0.
\end{eqnarray*}

The last relation implies
\begin{eqnarray}
\label{eq:40c}&&\nonumber
\mathbb{E}\left[ X_n(x,x',\beta;\bar{a}) 
Y''_n(x,x',\beta;\bar{a})\right]\\ &&\approx \mathbb{E}\left[ 
X_n(x,x',\beta;\bar{a}) Y'_n(x,x',\beta;\bar{a})\right],
\end{eqnarray}
provided that
\begin{equation}
\label{eq:40d}
\liminf_{n \to \infty}\frac{\mathbb{E}\left[ X_n(x,x',\beta;\bar{a}) 
Y'_n(x,x',\beta;\bar{a})\right]}{\int_0^1 \! \ud u \! \int_0^1\! \ud 
\tau \gamma_n(u,\tau)^2}>0.
\end{equation}
In general, if $a_n \geq b_n \geq 0$ are some sequences of real numbers,
such that  \[\lim_{n \to \infty} (a_n -b_n)=0\] and $\liminf_n 
b_n >0$, then $a_n \approx b_n$. Indeed, \[1\leq  \lim_{n \to \infty} 
\frac{a_n}{b_n}=1+\lim_{n \to \infty}\frac{a_n-b_n}{b_n}\leq 1+ 
\frac{\lim_{n } (a_n-b_n)}{\liminf b_n}=1.\]
Now, to prove the relation (\ref{eq:40d}) and therefore 
Eq.~(\ref{eq:40c}), one uses Eq.~(\ref{eq:12}) together with 
Eq.~(\ref{eq:40}) to show that the expression (\ref{eq:40d}) is 
greater or equal than
\begin{eqnarray*}&&
\frac{\sigma^4}{2} \mathbb{E}\bigg\{ X_\infty(x,x',\beta;\bar{a}) 
\int_0^1 \! \ud u \! \int_0^1\! \ud \tau \, g_2(u,\tau)\\&& \times 
{V}''[x_r(u)+\sigma B_{u}^0(\bar{a})]{V}''[x_r(\tau)+\sigma 
B^0_\tau(\bar{a})]\bigg\}.
\end{eqnarray*}
Because $g_2(u,\tau)$ was assumed to be strictly positive definite, 
the last quantity is zero only if ${V}''[x_r(u)+\sigma 
B_{u}^0(\bar{a})]=0$ for almost all $\bar{a}$, which only happens 
when the potential $V(x)$ is linear (i.e., its second order 
derivative is zero). Therefore, the relation (\ref{eq:40c}) holds for 
all non-linear potentials. However, the relation (\ref{eq:40c}) holds 
for linear but non-constant potentials too, because in this case
\[\mathbb{E}_n 
Y''_n(x,x',\beta;\bar{a})=\mathbb{E}_nY'_n(x,x',\beta;\bar{a}),\]
as shown be Eq.~(\ref{eq:40e}). Therefore, the signs $\lessapprox$ in 
Eq.~(\ref{eq:40b}) can be replaced by $\approx$.

We now use the positivity of the functions $Y'_n(x,x',\beta;\bar{a})$ 
and $Y''_n(x,x',\beta;\bar{a})$ and argue as in the proof of 
Eq.~(\ref{eq:31})  to conclude that
\begin{eqnarray*}
&&\nonumber
\mathbb{E}\left[ X_n(x,x',\beta;\bar{a}) 
Y'_n(x,x',\beta;\bar{a})\right]\\ &&\approx \mathbb{E}\left[ 
X_\infty(x,x',\beta;\bar{a}) Y'_n(x,x',\beta;\bar{a})\right]
\end{eqnarray*}
and
\begin{eqnarray*}
&&\nonumber
\mathbb{E}\left[ X_n(x,x',\beta;\bar{a}) 
Y''_n(x,x',\beta;\bar{a})\right]\\ &&\approx \mathbb{E}\left[ 
X_\infty(x,x',\beta;\bar{a}) Y''_n(x,x',\beta;\bar{a})\right],
\end{eqnarray*}
respectively.
Corroborating the last two relations with Eq.~(\ref{eq:40c}) and 
Eq.~(\ref{eq:40b}), we learn that
\begin{eqnarray}&&
\label{eq:40f} \nonumber
  \frac{\beta^2}{2}\mathbb{E}\left[ X_\infty(x,x',\beta;\bar{a}) 
Y'_n(x,x',\beta;\bar{a})\right]
\\&&\approx\rho(x,x';\beta)-\rho_n^{\text{PA}}(x,x';\beta)
  \\&& \approx \frac{\beta^2}{2}\mathbb{E}\left[ 
X_\infty(x,x',\beta;\bar{a}) 
Y''_n(x,x',\beta;\bar{a})\right].\nonumber
\end{eqnarray}

With the help of Eq.~(\ref{eq:40f}), we are ready to prove the 
following theorem
\begin{2}
\label{th:2}
Assume $V(x)$ is a Kato-class potential that lies in $\cap_\alpha 
W^{2,2}_\alpha (\mathbb{R})$. Then,
\begin{eqnarray}
\label{eq:42} \nonumber
\rho(x,x';\beta)-\rho_n^{\text{PA}}(x,x';\beta) \approx 
\frac{\beta^2}{2}\bigg\{  \int_0^1 \! \ud u \! \int_0^1\! \ud \tau 
\gamma_n(u,\tau)\\ \times K_{x,x'}^\beta(u,\tau)-\frac{1}{2}\left[ 
\int_0^1 \! \ud u \! \int_0^1\! \ud \tau 
\gamma_n(u,\tau)^2\right]\quad\\ \times\left[ \int_0^1 \! \ud u \! 
\int_0^1\! \ud \tau 
g_2(u,\tau)Q_{x,x'}^\beta(u,\tau)\right]\bigg\},\nonumber
\end{eqnarray}
where
\begin{eqnarray}\nonumber
\label{eq:43}
Q_{x,x'}^\beta(u,\tau)=
{\sigma^4} \mathbb{E}\Big\{ X_\infty(x,x',\beta;\bar{a}) 
{V}''[x_r(u)+\sigma B_{u}^0(\bar{a})]\\ \times {V}''[x_r(\tau)+\sigma 
B^0_\tau(\bar{a})]\Big\}.\quad
\end{eqnarray}
\end{2}

\emph{Proof} Let us notice that if $a_n \approx b_n$, then $a_n 
\approx 2b_n-a_n$. Indeed,
\[
\lim_{n\to \infty} \frac{2b_n-a_n}{a_n}= 2\lim_{n\to \infty} 
\frac{b_n}{a_n}-1 =1.
\] Using this observation together with the equation (\ref{eq:40f}), 
we deduce that
\begin{eqnarray}&&
\label{eq:43a} \nonumber
\rho(x,x';\beta)-\rho_n^{\text{PA}}(x,x';\beta)
  \\&& \approx \frac{\beta^2}{2}\mathbb{E}\left[ 
X_\infty(x,x',\beta;\bar{a}) Y_n(x,x',\beta;\bar{a})\right],
\end{eqnarray}
where
\[Y_n(x,x',\beta;\bar{a})=2Y''_n(x,x',\beta;\bar{a})-Y'_n(x,x',\beta;\bar{a}).\]
 From Eq.~(\ref{eq:40}) and Eq.~(\ref{eq:40a}) we learn that
\[Y_n(x,x',\beta;\bar{a})=T_n(x,x',\beta;\bar{a})-Z_n(x,x',\beta;\bar{a}),\]
where
\begin{eqnarray*}&&
Z_n(x,x',\beta;\bar{a})=\frac{\sigma^4}{2} \int_0^1 \! \ud u \! 
\int_0^1\! \ud \tau \, \gamma_n(u,\tau)^2  \\ &&\times 
\overline{V}_{u,n}''[x_r(u)+\sigma 
S_{u}^n(\bar{a})]\overline{V}_{\tau,n}''[x_r(\tau)+\sigma 
S^n_\tau(\bar{a})]\quad
\end{eqnarray*}
If we also introduce the random variable
\begin{eqnarray*}&&
Z'_n(x,x',\beta;\bar{a})=\frac{\sigma^4}{2}\left[ \int_0^1 \! \ud u 
\! \int_0^1\! \ud \tau \, \gamma_n(u,\tau)^2 \right]\\&& \bigg\{ 
\int_0^1 \! \ud u \! \int_0^1\! \ud \tau \, g_2(u,\tau) 
{V}''[x_r(u)+\sigma B_{u}^0(\bar{a})] \\ 
&&\times{V}''[x_r(\tau)+\sigma B^0_\tau(\bar{a})]\bigg\},\quad
\end{eqnarray*}
we notice that the right-hand side of  Eq.~(\ref{eq:42}) is precisely
\[\frac{\beta^2}{2}\mathbb{E}\left\{X_\infty(x,x',\beta;\bar{a})\left[T_n(x,x',\beta;\bar{a})-Z'_n(x,x',\beta;\bar{a})\right]\right\}.\]

Now, if the potential is linear but not constant, the theorem follows 
trivially because the functions $Z_n(x,x',\beta;\bar{a})$ and 
$Z'_n(x,x',\beta;\bar{a})$ are identically zero. Therefore, for the 
remainder of the proof, we assume that  the potential is not linear. 
Let us set
\[\Delta 
Z_n(x,x',\beta;\bar{a})=Z_n(x,x',\beta;\bar{a})-Z'_n(x,x',\beta;\bar{a})\]
and notice again that the theorem follows trivially from 
Eq.~(\ref{eq:43a}) provided that
\begin{equation}
\label{eq:43b}
\lim_{n \to \infty} 
\frac{\left|\mathbb{E}\left[X_\infty(x,x',\beta;\bar{a})\Delta 
Z_n(x,x',\beta;\bar{a})\right]\right|}{\mathbb{E}\left[X_\infty(x,x',\beta;\bar{a}) 
Y_n(x,x',\beta;\bar{a})\right]}=0.
\end{equation}
In order to prove that the limit in the last equation is indeed zero, 
we observe that
since $\mathbb{E}\,[X_\infty Y_n]\approx \mathbb{E}\,[X_\infty Y'_n]$ 
and since $Y'_n \geq Z_n$,
we have
\[
\lim_{n \to \infty} 
\frac{\mathbb{E}\left[X_\infty(x,x',\beta;\bar{a}) 
Y_n(x,x',\beta;\bar{a})\right]}{\mathbb{E}\left[X_\infty(x,x',\beta;\bar{a}) 
Z_n(x,x',\beta;\bar{a})\right]}\geq 1.
\]
Therefore, the limit in Eq.~(\ref{eq:43b}) is smaller or equal than
\begin{equation}
\label{eq:43c}
\lim_{n \to \infty} 
\frac{\left|\mathbb{E}\left[X_\infty(x,x',\beta;\bar{a})\Delta 
Z_n(x,x',\beta;\bar{a})\right]\right|}{\mathbb{E}\left[X_\infty(x,x',\beta;\bar{a}) 
Z_n(x,x',\beta;\bar{a})\right]}.
\end{equation}
However, the limit in Eq.~(\ref{eq:43c}) is zero because 
$\mathbb{E}\,[X_\infty Z_n]\approx \mathbb{E}\,[X_\infty Z'_n]$. This 
last statement follows from the equality
\begin{eqnarray*}
\lim_{n \to \infty} 
\frac{\mathbb{E}\left[X_\infty(x,x',\beta;\bar{a}) 
Z_n(x,x',\beta;\bar{a})\right]}{ \int_0^1 \! \ud u \! \int_0^1\! \ud 
\tau \, \gamma_n(u,\tau)^2 }\\=
\lim_{n \to \infty} 
\frac{\mathbb{E}\left[X_\infty(x,x',\beta;\bar{a}) 
Z'_n(x,x',\beta;\bar{a})\right]}{\int_0^1 \! \ud u \! \int_0^1\! \ud 
\tau \, \gamma_n(u,\tau)^2 }\\= \frac{1}{2}\int_0^1 \! \ud u \! 
\int_0^1\! \ud \tau \, g_2(u,\tau)Q^\beta_{x,x'}(u,\tau),
\end{eqnarray*}
because the above common limit is finite and non-zero by an argument 
similar to the one employed in the proof of Theorem~1. The proof of 
Theorem~2 is therefore concluded. $\quad \Box$

The kernel $Q_{x,x'}^\beta(u,\tau)$ shares most of the properties 
proved for $K_{x,x'}^\beta(u,\tau)$. In particular, 
$Q_{x,x'}^\beta(u,\tau)$ is symmetric at the permutation of the 
variables $u$ and $\tau$ and, if $\tau \leq u$,
\begin{eqnarray}\nonumber
\label{eq:44}
Q_{x,x'}^\beta(u,\tau)=\sigma^4\int_{\mathbb{R}}\int_{\mathbb{R}} 
\rho(x,y;\tau\beta)\rho[y,z;(u-\tau)\beta] \\ \times\rho[z,x';(1-u)\beta] 
V''(y)V''(z)\,\ud y\, \ud z.
\end{eqnarray}
In addition, $Q_{x,x'}^\beta(u,\tau)$ is continuous in the variables 
$\tau$ and $u$
and
\begin{equation}
\label{eq:45}
Q_{x,x'}^\beta(1-u,1-\tau)=Q_{x',x}^\beta(u,\tau).
\end{equation}

The equation (\ref{eq:42}) remains true for multidimensional systems 
provided that we define the kernel $K_{x,x'}^\beta(u,\tau)$ as in 
Eq.~(\ref{eq:38}), while for the kernel $Q_{x,x'}^\beta(u,\tau)$ we 
employ the formula
\begin{eqnarray}\nonumber
\label{eq:46}
Q_{x,x'}^\beta(u,\tau)=\int_{\mathbb{R}^d}\ud y\int_{\mathbb{R}^d}\ud 
z\, \rho(x,y;\tau\beta)\rho[y,z;(u-\tau)\beta] \\ \times 
\rho[z,x';(1-u)\beta]\left\{\sum_{i,j}^d 
\sigma_i^2\sigma_j^2\left[\partial^2_{i,j}V(y)\,\partial^2_{i,j}V(z)\right]\right\}. 
\quad
\end{eqnarray}

\section{Convergence of the PA-WFPI method}
As argued in Ref.~(\onlinecite{Pre02}), among all possible series 
representations, the Wiener-Fourier one is special in the sense that 
it has the fastest asymptotic rate of convergence. For this reason, 
it is of interest to establish what the asymptotic rates of 
convergence are for different classes of potentials.  In this 
section, we shall show that for potentials lying in $\cap_\alpha 
W^{1,2}_\alpha (\mathbb{R})$, the asymptotic rate of convergence is 
always better than ${O}(1/n^2)$, while for potentials lying 
in $\cap_\alpha W^{2,2}_\alpha (\mathbb{R})$, the asymptotic rate of 
convergence is ${O}(1/n^3)$. For the latter case, we also 
establish the exact convergence constant.

\subsection{Potentials with first order Sobolev derivatives}

We begin the convergence analysis with the Kato-class potentials that 
lie in $\cap_\alpha W^{1,2}_\alpha (\mathbb{R})$. For the 
Wiener-Fourier basis, we have
\begin{equation}
\label{eq:47}
\gamma_n(u,\tau)=\frac{2}{\pi^2}\sum_{k=n+1}^{\infty}\frac{\sin(k\pi 
u)\sin(k\pi \tau)}{k^2}.
\end{equation}
If $f(t)$ is a square integrable function on the interval $[0,1]$, 
then we have the estimate
\begin{eqnarray}
\label{eq:48}\nonumber &&
\int_0^1 \int_0^1 \gamma_n(u,\tau) f(u)f(t)\ud u \ud \tau \\&&= 
\frac{1}{\pi^2}\sum_{k=n+1}^{\infty}\frac{1}{k^2}\left[\sqrt{2}\int_0^1 
f(u) \sin(k\pi u)\ud u\right]^2 \\&& \leq \frac{1}{\pi^2 
(n+1)^2}\sum_{k=n+1}^{\infty}\left[\sqrt{2}\int_0^1 f(u) \sin(k\pi 
u)\ud u\right]^2. \nonumber
\end{eqnarray}
Now, the first part of Theorem~1 allows us to write
\begin{eqnarray}
\label{eq:49} \nonumber
\rho(x,x';\beta)-\rho_n^{\text{PA}}(x,x';\beta) \lessapprox 
\frac{\beta^2\sigma^2}{2} 
\mathbb{E}\bigg\{X_{\infty}(x,x',\beta;\bar{a})\\\times\int_0^1 \! 
\ud u \! \int_0^1\! \ud \tau  \gamma_n(u,\tau) V'[x_r(u)+\sigma 
B_{u}^0(\bar{a})]\\  \times V'[x_r(\tau)+\sigma 
B_{\tau}^0(\bar{a})]\bigg\} \nonumber
\end{eqnarray}
and using Eq.~(\ref{eq:48}), we obtain
\begin{eqnarray}
\label{eq:50} \nonumber
\lim_{n\to 
\infty}(n+1)^2\left[\rho(x,x';\beta)-\rho_n^{\text{PA}}(x,x';\beta)\right] 
\leq \frac{\beta^2\sigma^2}{2\pi^2}  \\\times \lim_{n\to \infty} 
\mathbb{E}\left[X_{\infty}(x,x',\beta;\bar{a})Z_n(x,x',\beta;\bar{a})\right],
\end{eqnarray}
where
\begin{eqnarray*}&&
Z_n(x,x',\beta;\bar{a})\\&&=\sum_{k=n+1}^\infty 
\left\{\int_0^1\sqrt{2}\sin(k\pi u) V'[x_r(u)+\sigma 
B_{u}^0(\bar{a})]\ud u\right\}^2.
\end{eqnarray*}
We notice that the sequence of functions $Z_n(x,x',\beta;\bar{a})$ is 
monotonically decreasing and convergent to zero. Moreover, since the 
functions $\{\sqrt{2}\sin(k\pi u)\}_{k\geq 1}$ form an orthonormal 
system, the Bessel inequality implies that
\begin{eqnarray*}&&
Z_0(x,x',\beta;\bar{a})\leq \int_0^1  V'[x_r(u)+\sigma 
B_{u}^0(\bar{a})]^2 \ud u.
\end{eqnarray*}
We then have the inequality
\begin{eqnarray*}&&
\mathbb{E}\left[X_{\infty}(x,x',\beta;\bar{a})Z_0(x,x',\beta;\bar{a})\right]\\&& 
\leq
\int_0^1\ud u\int_{\mathbb{R}} \ud y 
\rho(x,y;u\beta)\rho[y,x';(1-u)\beta]V'(y)^2  < \infty.
\end{eqnarray*}
The last integral is finite because $V(x) \in \cap_\alpha 
W^{1,2}_\alpha (\mathbb{R})$.

In these conditions, the dominated convergence theorem\cite{Fol99} shows that
\[
\lim_{n\to \infty}  \mathbb{E}\left[X_{\infty}(x,x',\beta;\bar{a})Z_n(x,x',\beta;\bar{a})\right]=0,
\]
and we have just proved the following theorem
\begin{3}
If $V(x)$ is a Kato-class potential that lies in $\cap_\alpha 
W^{1,2}_\alpha (\mathbb{R})$ and the basis $\{\lambda_k(u)\}_{k\geq 
1}$ is the Wiener-Fourier basis, then
\begin{equation}
\label{eq:51}
\lim_{n\to 
\infty}(n+1)^2\left[\rho(x,x';\beta)-\rho_n^{\text{PA}}(x,x';\beta)\right]=0.
\end{equation}
\end{3}

\emph{Observation} The previous theorem says that the convergence is 
always better than ${O}(1/n^2)$ [i.e., it is $o(1/n^2)$], but it does not say how much 
better. If we are interested in computing an absolute asymptotic 
bound, then the bound given by the relation
\begin{eqnarray}
\label{eq:52} \nonumber &&
\rho(x,x';\beta)-\rho_n^{\text{PA}}(x,x';\beta) \lessapprox 
\frac{\beta^2\sigma^2}{2\pi^2(n+1)^2}  \\&&\times 
\mathbb{E}\left[X_{\infty}(x,x',\beta;\bar{a})Z_0(x,x',\beta;\bar{a})\right] 
\leq
\frac{\hbar^2\beta^3}{2\pi^2m_0(n+1)^2}\qquad\\&&\times \int_0^1 \ud 
\theta\int_{\mathbb{R}} \ud y\, 
\rho(x,y;\theta\beta)V'(y)^2\rho[y,x';(1-\theta)\beta]  \nonumber
\end{eqnarray}
is the only choice if no additional information about the potential 
$V(x)$ is available. Theorem~3 remains true as stated for 
multidimensional systems, while in an even more compact notation, the 
estimate given by Eq.~(\ref{eq:52}) becomes
\begin{eqnarray}
\label{eq:53} \nonumber &&
\rho(x,x';\beta)-\rho_n^{\text{PA}}(x,x';\beta) \lessapprox 
\frac{\hbar^2\beta^3}{2\pi^2(n+1)^2}  \\&&\times  \int_0^1 
\left\langle x\left|e^{-\theta \beta H}\left[\sum_{i=1}^d 
\frac{(\partial_i V)^2}{m_{0,i}} \right] e^{-(1-\theta) \beta 
H}\right|x' \right\rangle \ud \theta. \qquad
\end{eqnarray}

\subsection{Potentials having second order Sobolev derivatives}

As we have shown in Section~III.B, all series representations attain 
their fastest asymptotic convergence on the class of potentials 
$\cap_\alpha W^{2,2}_\alpha (\mathbb{R})$. Of course, this is also 
true of the Wiener-Fourier series. Moreover, since this series is 
optimal as asymptotic convergence, the rates of convergence 
established in this section set a limit on the asymptotic behavior of 
the partial averaging method. In this section, we prove that the 
asymptotic convergence of the PA-WFPI sequence of approximations to 
the density matrix is ${O}(1/n^3)$ and we establish the 
corresponding convergence constant.

We start our proof by analyzing the first term in Eq.~(\ref{eq:42}).
Let us notice that   $\gamma_n(1-u,1-\tau)=\gamma_n(u,\tau)$ because 
$\sin[k\pi (1-t)]=-(-1)^k\sin(k\pi t)$.
By using this  relation together with Eq.~(\ref{eq:37}), one computes
\begin{eqnarray*}
\int_0^1 \! \ud u \! \int_u^1\! \ud \tau \, 
g_n(u,\tau)K_{x,x'}^{\beta}(u,\tau)\\=\int_0^1 \! \ud u \! 
\int_0^{1-u}\! \ud \tau \, 
g_n(u,1-\tau)K_{x,x'}^{\beta}(u,1-\tau)\\=\int_0^1 \! \ud u \! 
\int_0^{u}\! \ud \tau \, 
g_n(1-u,1-\tau)K_{x,x'}^{\beta}(1-u,1-\tau)\\= \int_0^1 \! \ud u \! 
\int_0^{u}\! \ud \tau \, g_n(u,\tau)K_{x',x}^{\beta}(u,\tau).
\end{eqnarray*}
A little thought shows that the first term in Eq.~(\ref{eq:42}) takes the form
\begin{equation}
\label{eq:54}
  \frac{\beta^2}{\pi^2} \sum_{k=n+1}^{\infty}\int_0^1 \! \ud u \! 
\int_0^u\! \ud \tau \, \frac{\sin(k\pi u)\sin(k\pi \tau)} 
{k^2}K(u,\tau),
\end{equation}
where \[K(u,\tau)=K_{x,x'}^{\beta}(u,\tau)+K_{x',x}^{\beta}(\tau,u).\]

The asymptotic rate of convergence of this first term is dictated by 
the decay of the terms of the series appearing in Eq.~(\ref{eq:54}). 
By integration by parts, we shall later show that these terms are of 
the form
${A_k}/{k^4}+{B_k}/{k^4} $, where $\{B_k\}_{k\geq 1}$ is a sequence 
convergent to zero. Because of this property, the terms $B_k/k^4$ do 
not contribute to the final asymptotic convergence, at least as far 
as the  ${O}(1/n^3)$ is concerned. Indeed, we have
\begin{eqnarray*}
\left|\lim_{n\to \infty} n^3 \sum_{k=n+1}^\infty 
{B_k}/{k^4}\right|\leq \lim_{n\to \infty} n^3\sum_{k=n+1}^\infty 
{|B_k|}/{k^4}\\ \leq \lim_{n\to \infty} \left(\sup_{k\geq 
n}|B_k|\right) \sum_{k=n+1}^\infty {n^3}/{k^4}\\ \leq 
\frac{1}{3}\lim_{n\to \infty}\left(\sup_{k\geq 
n}|B_k|\right)=0,\end{eqnarray*}
which proves our assertion. In the above, we used the inequality
\[\sum_{k=n+1}^\infty {n^3}/{k^4}\leq n^3 \int_{n}^{\infty}x^{-4} \ud 
x =\frac{1}{3}\] as well as the fact that $B_k \to 0$ implies
\[\limsup |B_k| = \lim_{n\to \infty}\left(\sup_{k\geq n}|B_k|\right)=0. \]
Now, to readily identify which of the various expressions make up the 
terms $B_k$ that decay to zero, the reader may employ the 
Riemann-Lebesgue lemma,\cite{Gra00} which in our case states that if 
$f(t)$ is integrable, then
\[
\lim_{k\to \infty} \int_0^1 \cos(k\pi t) f(t)\ud t =\lim_{k\to 
\infty} \int_0^1 \sin(k\pi t) f(t)\ud t =0.
\]

We start with the equation (\ref{eq:54}) and by  integration by parts 
against the variable $\tau$, we obtain
\begin{eqnarray}
\label{eq:55}&&
\frac{1}{k^2}\int_0^1\ud u \sin(k\pi u)\int_0^u \sin(k \pi 
\tau)K(u,\tau)\ud \tau=\nonumber \\&& \int_0^1  \sin(k \pi u) 
\frac{K(u,0)-\cos(k\pi u) K(u,u)}{k^3\pi} \ud u\\&& \nonumber + 
\frac{1}{k^3\pi}\int_0^1 \ud u \sin(k\pi u) \int_0^u \cos(k\pi 
\tau)\frac{\partial}{\partial \tau}K(u,\tau)\ud \tau. \quad
\end{eqnarray}
Let us focus on the first term of the relation (\ref{eq:55}). By 
integration by parts against $u$ we have
\begin{eqnarray}
\label{eq:56}
\frac{1}{k^3\pi}\int_0^1 \sin(k \pi u)K(u,0)\ud u= 
\frac{K(0,0)-(-1)^kK(1,0)}{k^4\pi^2}\nonumber \\ 
+\frac{1}{k^4\pi^2}\int_0^1\cos(k\pi u) \frac{\partial}{\partial 
u}K(u,0) \ud u \quad
\end{eqnarray}
and
\begin{eqnarray}
\label{eq:57}&&
\frac{1}{k^3\pi}\int_0^1 \sin(k \pi u)\cos(k \pi u)K(u,u)\ud u 
=\frac{1}{2k^3\pi}\nonumber \\&& \times \int_0^1 \sin(2k \pi 
u)K(u,u)\ud u =\frac{K(0,0)-K(1,1)}{4k^4\pi^2}\quad \\&& 
\nonumber+\frac{1}{4k^4\pi^2} \int_0^1\cos(2k\pi u) 
\frac{\partial}{\partial u}K(u,u) \ud u,
\end{eqnarray}
respectively. As argued before by means of the Riemann-Lebesgue 
lemma, the last terms in the relations (\ref{eq:56}) and 
(\ref{eq:57}) do not contribute to the asymptotic rates of 
convergence and therefore, they will be dropped. Moreover, even the 
term
$K(1,0){(-1)^k }/{k^4 \pi^2}$ does not contribute to the asymptotic 
rate of convergence because
\begin{eqnarray*}
\left|\sum_{k=n+1}^\infty \frac{(-1)^k}{k^4}\right|= 
\sum_{k=0}^\infty\left[ 
\frac{1}{(n+2k+1)^4}-\frac{1}{(n+2k+2)^4}\right]\\ = 
\sum_{k=0}^\infty \int_{n+2k+1}^{n+2k+2}4x^{-5} \ud x \leq 4 
\int_{n+1}^\infty x^{-5}\ud x= \frac{1}{(n+1)^4}
\end{eqnarray*}
decays faster than $1/n^3$. Finally, the first term in 
Eq.~(\ref{eq:57}) is zero because
\begin{equation}
\label{eq:58}
K(0,0)=K(1,1)=\sigma^2\rho(x,x';\beta)\left[V'(x)^2+V'(x')^2\right].
\end{equation}

Now, we go back to the formula (\ref{eq:55}) and integrate by parts 
the last term against the variable $u$. One obtains
\begin{eqnarray}
\label{eq:59}
  \frac{1}{k^3\pi}\int_0^1 \ud u \sin(k\pi u) \int_0^u \cos(k\pi 
\tau)\frac{\partial}{\partial \tau}K(u,\tau)\ud \tau \nonumber \\ 
\nonumber = \frac{-(-1)^k}{k^4\pi^2}\int_0^1\cos(k\pi 
\tau)\frac{\partial}{\partial \tau}K(1,\tau)\ud \tau+ 
\frac{1}{k^4\pi^2}\\ \times\int_0^1 \cos(k\pi u)^2 
\frac{\partial}{\partial \tau}K(u,\tau)\bigg|_{\tau=u}\ud u + 
\frac{1}{k^4\pi^2}\\ \nonumber \times \int_0^1 \ud u \cos(k\pi u) 
\int_0^u \cos(k\pi \tau)\frac{\partial^2}{\partial \tau\partial 
u}K(u,\tau)\ud \tau.
\end{eqnarray}
Again, the first and the last terms from the above expansion do not 
contribute to the asymptotic rate of convergence by the 
Riemann-Lebesgue lemma. For the last term, notice that by symmetry
\begin{eqnarray*}
\int_0^1 \ud u \cos(k\pi u) \int_0^u \cos(k\pi 
\tau)\frac{\partial^2}{\partial \tau\partial u}K(u,\tau)\ud \tau\\ 
=\frac{1}{2}\int_0^1 \int_0^1\cos(k\pi u)  \cos(k\pi 
\tau)\frac{\partial^2}{\partial \tau\partial u}K(u,\tau)\ud u\ud \tau
\end{eqnarray*}
and then the bidimensional version of the Riemann-Lebesgue lemma 
applies. Therefore, the contribution of Eq.~(\ref{eq:59}) to the 
asymptotic rate of convergence is due solely to the term
\begin{eqnarray}
\label{eq:60}&&
\frac{1}{k^4\pi^2}\int_0^1 \cos(k\pi u)^2 \frac{\partial}{\partial 
\tau}K(u,\tau)\bigg|_{\tau=u}\ud u  \nonumber\\&& = \frac{1}{2k^4\pi^2} 
\int_0^1  \frac{\partial}{\partial \tau}K(u,\tau)\bigg|_{\tau=u}\ud u 
+\frac{1}{2k^4\pi^2}\\&& \times \int_0^1 \cos(2k\pi u) 
\frac{\partial}{\partial \tau}K(u,\tau)\bigg|_{\tau=u}\ud u \nonumber
\end{eqnarray}
Yet again, the last term of Eq.~(\ref{eq:60}) does not contribute to 
the asymptotic rate  by the same Riemann-Lebesgue lemma.

Combining everything by means of Eq.~(\ref{eq:55}), we end up with 
the following asymptotic expression for the relation (\ref{eq:54}):
\begin{equation}
\label{eq:61}
\frac{\beta^2}{3\pi^4 n^3} \left[K(0,0)+\frac{1}{2}\int_0^1 
\frac{\partial}{\partial \tau}K(u,\tau)\bigg|_{\tau=u}\ud u \right],
\end{equation}
where we also used the equality
\[ \lim_{n\to \infty}\left( n^3 \sum_{k=n+1}^\infty 
\frac{1}{k^4}\right)=\frac{1}{3}.\]
This last equality can be deduced  by letting $n\to \infty$ in 
Eq.~(\ref{eq:B2a}) of Appendix~B.

The exact expression for the term $K(0,0)$ is given by 
Eq.~(\ref{eq:58}). To evaluate the term
\[
\int_0^1  \frac{\partial}{\partial \tau}K(u,\tau)\bigg|_{\tau=u}\ud u,
\] we need to remember that the density matrix satisfies the Bloch equation
\[
-\frac{\hbar^2}{2 m_0}\frac{\partial^2}{\partial {x'}^2} 
\rho(x,x';\beta)+V(x')\rho(x,x';\beta)=-\frac{\partial}{\partial 
\beta}\rho(x,x';\beta).
\]
Using the Bloch equation, one can justify by explicit calculation the 
following equality:
\begin{eqnarray*}&&
\frac{\partial}{\partial \tau}K_{x,x'}^\beta(u,\tau)=\frac{\hbar^2 
\sigma^2\beta}{2m_0} \int_{\mathbb{R}}\int_{\mathbb{R}}\bigg\{ 
\left[\frac{\partial^2}{\partial y^2} \rho(x,y;\tau\beta)\right]\\&& 
\times \rho[y,z;(u-\tau)\beta]- 
\rho(x,y;\tau\beta)\frac{\partial^2}{\partial 
y^2}\rho[y,z;(u-\tau)\beta]\bigg\} \\&& \times\rho[z,x';(1-u)\beta] 
V'(y)V'(z)\,\ud y\, \ud z.
\end{eqnarray*}
and by repeated integration by parts against $y$,
\begin{eqnarray*}&&
\frac{\partial}{\partial 
\tau}K_{x,x'}^\beta(u,\tau)=-\frac{\hbar^2\sigma^2\beta}{2m_0} 
\int_{\mathbb{R}}\int_{\mathbb{R}} \bigg[2\frac{\partial}{\partial y} 
\rho(x,y;\tau\beta) V''(y)\\&&+ \rho(x,y;\tau\beta) V'''(y)\bigg] 
\rho[y,z;(u-\tau)\beta]\\ &&\times\rho[z,x';(1-u)\beta] V'(z)\,\ud 
y\, \ud z.
\end{eqnarray*}
The third order derivative $V'''(x)$ is to be interpreted as a 
distributional derivative.
Again in the sense of distributions, as $\tau \to u$, we have 
$\rho[y,z;(u-\tau)\beta]\to \delta(z-y)$ and therefore,
\begin{eqnarray*}&&
\frac{\partial}{\partial 
\tau}K_{x,x'}^\beta(u,\tau)\bigg|_{\tau=u}=-\frac{\hbar^2\sigma^2\beta}{2m_0} 
\int_{\mathbb{R}} \bigg[2\frac{\partial}{\partial y} \rho(x,y;u\beta) 
V''(y)\\&&+ \rho(x,y;u\beta) V'''(y)\bigg] \rho[y,x';(1-u)\beta] 
V'(y)\,\ud y.
\end{eqnarray*}
By integration by parts of the term containing the third derivative 
$V'''(y)$, we obtain
\begin{eqnarray*}
\frac{\partial}{\partial 
\tau}K_{x,x'}^\beta(u,\tau)\bigg|_{\tau=u}=\frac{\hbar^2\sigma^2\beta}{2m_0} 
\int_{\mathbb{R}}\rho(x,y;u\beta)\rho[y,x';(1-u)\beta]\\ \times 
V''(y)^2 \ud y-\frac{\hbar^2\sigma^2\beta}{2m_0}\int_{\mathbb{R}} 
V'(y)V''(y) \bigg\{\left[\frac{\partial}{\partial y} 
\rho(x,y;u\beta)\right]\\ \times\rho[y,x';(1-u)\beta]- 
\rho(x,y;u\beta)\frac{\partial}{\partial 
y}\rho[y,x';(1-u)\beta]\bigg\}\ud y.
\end{eqnarray*}
A similar estimate holds for
\[\frac{\partial}{\partial \tau}K_{x,x'}^\beta(u,\tau)\bigg|_{\tau=u},\]
and can by obtained simply by permuting $x$ and $x'$.
The equality
\begin{eqnarray*}
\int_0^1 \!\int_{\mathbb{R}} 
V'(y)V''(y)\left[\frac{\partial}{\partial y} 
\rho(x,y;u\beta)\right]\\ \times\rho[y,x';(1-u)\beta] \ud u=\int_0^1 
\!\int_{\mathbb{R}} V'(y)V''(y)\\\times 
\rho(x',y;u\beta)\frac{\partial}{\partial y} \rho[y,x;(1-u)\beta] \ud 
u
\end{eqnarray*}
as well as the one obtained by permuting $x$ and $x'$ are readily 
obtained by use of the substitution $u'=1-u$ and by use of the 
symmetry of  the density matrices. We leave it for the reader to 
utilize these equalities and verify that
\begin{eqnarray}
\label{eq:62}&&
\int_0^1 \frac{\partial}{\partial \tau}K(u,\tau)\bigg|_{\tau=u} \ud u 
=\int_0^1 \frac{\partial}{\partial 
\tau}K_{x,x'}^\beta(u,\tau)\bigg|_{\tau=u} \ud u \nonumber \\&& + 
\int_0^1 \frac{\partial}{\partial 
\tau}K_{x',x}^\beta(u,\tau)\bigg|_{\tau=u} \ud u = 
\frac{\hbar^2\sigma^2\beta}{m_0}\\&& \times \nonumber 
\int_0^1\int_{\mathbb{R}}\rho(x,y;u\beta)\rho[y,x';(1-u)\beta] 
V''(y)^2 \ud y \ud u.
\end{eqnarray}

Replacing Eq.~(\ref{eq:62}) in Eq.~(\ref{eq:61}),  one ends up with 
the following expression for the first term of the 
equation~(\ref{eq:42}):
\begin{eqnarray}
\label{eq:63}\nonumber
  \frac{\hbar^2\beta^3}{3\pi^4 m_0n^3} 
\bigg\{\rho(x,x';\beta)\left[V'(x)^2+V'(x')^2\right]+\frac{\hbar^2\beta}{2m_0} 
\\\times  \int_0^1\left\langle x\Big|e^{-\beta \theta H} V''^2 
e^{-\beta (1-\theta) H}\Big|x'\right\rangle \ud \theta \bigg\}.
\end{eqnarray}

The second term in Eq.~(\ref{eq:42}) is a little easier to analyze. 
Eq.~(\ref{eq:B4}) from Appendix~B shows that for the Wiener-Fourier 
series,
\[
\lim_{n \to \infty} n^3 g_n(u,\tau) = \frac{1}{3\pi^4}\delta(u-\tau)
\]
in the sense of distributions. Using this equality and the relation 
(\ref{eq:43}), we learn that the decay of the last term of equation 
(\ref{eq:42}) is
\begin{eqnarray}\nonumber&&
\label{eq:64}\frac{\beta^2}{12\pi^4n^3}\int_0^1 Q_{x,x'}^\beta(u,u)\ud u \\&&=
\frac{\beta^2\sigma^4}{12\pi^4n^3}\int_0^1\left\langle 
x\Big|e^{-\beta \theta H} V''^2 e^{-\beta (1-\theta) 
H}\Big|x'\right\rangle \ud \theta \quad
\end{eqnarray}

Replacing Eq.~(\ref{eq:63}) and Eq.~(\ref{eq:64}) in 
Eq.~(\ref{eq:42}), one ends up with the following theorem
\begin{3}
\begin{eqnarray}
\label{eq:65}\nonumber &&
\lim_{n \to \infty} n^3 
\left[\rho(x,x';\beta)-\rho_n^{\text{PA}}(x,x';\beta)\right]
\\&& = \frac{\hbar^2\beta^3}{3\pi^4 m_0} 
\rho(x,x';\beta)\left[V'(x)^2+V'(x')^2\right]\\&&+\frac{\hbar^4\beta^4}{12\pi^4m_0^2} 
\int_0^1\left\langle x\Big|e^{-\beta \theta H} V''^2 e^{-\beta 
(1-\theta) H}\Big|x'\right\rangle \ud \theta . \nonumber
\end{eqnarray}
\end{3}
The $d$-dimensional analog can be deduced in a similar way by 
starting with Eqs.~(\ref{eq:38}) and (\ref{eq:46}) and by using the 
$d$-dimensional Bloch equation. It has the expression
\begin{widetext}
\begin{eqnarray}
\label{eq:66}\nonumber
\lim_{n \to \infty} n^3 
\left[\rho(x,x';\beta)-\rho_n^{\text{PA}}(x,x';\beta)\right]= 
\frac{\hbar^2\beta^3}{3\pi^4} 
\rho(x,x';\beta)\left\{\sum_{i=1}^d\frac{[\partial_iV(x)]^2+[\partial_iV(x')]^2}{m_{0,i}}\right\}
\\+\frac{\hbar^4\beta^4}{12\pi^4}  \int_0^1\left\langle 
x\left|e^{-\beta \theta H} \sum_{i,j=1}^d\frac{\left(\partial^2_{i,j} 
V\right)^2 }{m_{0,i}m_{0,j}}e^{-\beta (1-\theta) 
H}\right|x'\right\rangle \ud \theta .
\end{eqnarray}
\end{widetext}

We conclude this section by numerically verifying the findings of 
Theorem~4 for the simple case of an harmonic oscillator. In fact, we 
shall verify the following corollary, which, by trace invariance, is 
a direct consequence of Theorem~4
\begin{6}
\[
\lim_{n \to \infty} n^3 
\frac{Z(\beta)-Z^{PA}_n(\beta)}{Z(\beta)}=\frac{\int_{\mathbb{R}} 
\rho(x;\beta)\Delta Z(x;\beta)\ud x}{\int_{\mathbb{R}} 
\rho(x;\beta)\ud x},
\]
where
\[\Delta Z(x;\beta)=\frac{2\hbar^2\beta^3}{3\pi^4 m_0} 
V'(x)^2+\frac{\hbar^4\beta^4}{12\pi^4m_0^2} V''(x)^2.\]
\end{6}
The Corollary~1 gives an estimate for the relative error of the 
partition function as an average of a suitable estimating function. 
In practice, such an average can be evaluated by Monte Carlo 
integration if so desired.

In the remainder of this section, we shall verify the Corollary~1 for 
the simple case of the harmonic oscillator $V(x)=m_0\omega^2x^2/2$. 
The computations are performed in atomic units for a particle of mass 
$m_0=1$ and for the frequency $\omega=1$. The inverse temperature is 
set to $\beta=10$.

Since the density matrix for an harmonic oscillator is analytically 
known,\cite{Fey94} the convergence constant for the relative error of 
the partition function can be evaluated directly from Corollary~1 to 
be
\[
c_{PA}=\frac{\int_{\mathbb{R}} \rho(x;\beta)\Delta Z(x;\beta)\ud 
x}{\int_{\mathbb{R}} \rho(x;\beta)\ud x}=11.98.
\]
We can also evaluate it by studying the limit of the sequence
\[
c^{PA}_n=n^3 \frac{Z(\beta)-Z^{PA}_n(\beta)}{Z(\beta)}.
\]
In this respect, we mention that the terms $Z^{PA}_n(\beta)$ were 
previously evaluated in Appendix~B of Ref.~(\onlinecite{Pre02}). 
Also, because the even and the odd sequences $c^{PA}_{2n}$ and 
$c^{PA}_{2n+1}$ have a slightly different asymptotic behavior that 
generates an oscillatory pattern in plots, we choose to  plot them 
separately.

\begin{figure}[!tbp]
   \includegraphics[angle=270,width=8.5cm,clip=t]{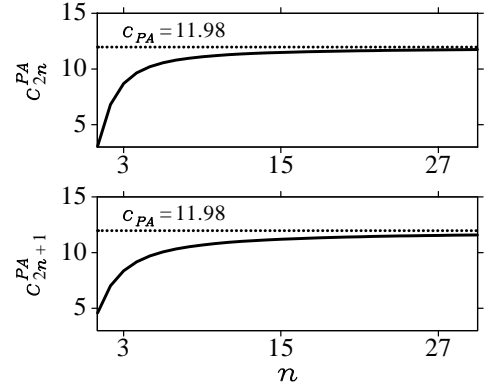}
%    \BoxedEPSF{Potential.eps scaled 300}
  \caption[sqr]
{\label{Fig:1}
Both the even sequence  $c^{PA}_{2n}$ and the odd sequence 
$c^{PA}_{2n+1}$ are seen  to converge to  $c_{PA}=11.98$, which is 
the value predicted by Corollary~1.
}
\end{figure}

In these conditions, the Corollary~1 says that
\[
c_{PA}= \lim_{n \to \infty} c^{PA}_n
\]
and the prediction is well verified for the harmonic oscillator, as 
Fig~\ref{Fig:1} shows. We regard this numerical example as strong 
evidence that the analysis performed in the present paper is correct. 
Indeed, it is hard to believe that the exact agreement between theory 
and numerical analysis for the harmonic oscillator is accidental.

\section{Summary and Discussion}
In this paper, we have performed a complete analysis of the
asymptotic rates of convergence of the partial averaging sequence of
approximations for the density matrix.  We have found that there are two
natural classes of potentials on which different series may achieve
their fastest convergence. If the two-point correlation
function
$\gamma_n(u,\tau)$ for the tail series is positive, then the natural
class is $\cap_\alpha W^{1,2}_\alpha (\mathbb{R})$ as shown by Theorem~1.
More generally,  all series representations achieve their fastest
convergence on the space of potentials $\cap_\alpha W^{2,2}_\alpha
(\mathbb{R})$. This is the statement of Theorem~2. Moreover, in both
cases, the cited theorems provide the exact asymptotic rates of
convergence for arbitrary series.

In Section~IV, we analyzed the special case of the Wiener-Fourier
series representation. Theorem~3 asserts that the asymptotic convergence
for this series is  $o(1/n^2)$ on the
Kato-class potentials that lie in $\cap_\alpha W^{1,2}_\alpha
(\mathbb{R})$. Moreover, the equations (\ref{eq:52}) and (\ref{eq:53})
provide useful (though conservative) bounds that control the
absolute asymptotic
error. The Wiener-Fourier series achieves its fastest asymptotic
convergence on the Kato-class potentials that lie in
$\cap_\alpha W^{2,2}_\alpha (\mathbb{R})$. This convergence is
${O}(1/n^3)$, with a convergence constant given by
Eq.~(\ref{eq:65}) for monodimensional systems and by Eq.~(\ref{eq:66})
for multidimensional systems, respectively.

The present results provide a relatively complete characterization
of the asymptotic convergence characteristics of the partial
averaging methods.  Beyond establishing sharp estimates of the
convergence constants for applications where the potential energy
has second-order derivatives, it should be possible to utilize the present
methods to characterize the asymptotic behavior of other
properties.  In particular, we would speculate that it should be
possible to obtain convergence constants for both the H-method and
T-method energy estimators using arguments similar to those developed in
the present paper. Moreover, we expect that the theorems proved in this paper will play a crucial role in establishing the asymptotic rates of convergence for the reweighted method, a technique related to the partial averaging method.

\begin{acknowledgments}
The authors acknowledge support from the National Science Foundation through
awards CHE-0095053 and CHE-0131114.
\end{acknowledgments}

\appendix
\section{}
In this appendix, we compute $\mathbb{E}(B_u^0)^2$, 
$\mathbb{E}(B_\tau^0)^2$, and $\mathbb{E}(B_u^0B_\tau^0)$. We begin 
with the last expression. We have
\begin{eqnarray}
\label{eq:A1}
\nonumber
\mathbb{E}(B_u^0B_\tau^0)=\sum_{k=1}^\infty 
\Lambda_k(u)\Lambda_k(\tau)\\=\int_0^u \int_0^\tau \sum_{k=1}^\infty 
\lambda_k(u')\lambda_k(\tau') \ud \tau' \ud u'\\=\int_0^u \int_0^\tau 
\left[\delta(u'-\tau')-1\right] \ud \tau' \ud u'\nonumber
\end{eqnarray}
where we used the completeness relation. In this sense, the reader 
should remember that $\{\lambda_k(u)\}_{k\geq 1}$ together with the 
constant function make up a complete basis in $L^2([0,1])$.
Now, let $I_u(u')$ and $I_\tau(\tau')$ denote the indicator functions 
of the intervals $[0,u]$ and $[0,\tau]$, respectively. The last 
integral in Eq.~(\ref{eq:A1}) becomes
\begin{eqnarray}
\label{eq:A2}\nonumber
\mathbb{E}(B_u^0B_\tau^0)=\int_0^1 \int_0^1 I_u(u')I_\tau(\tau') 
\delta(u'-\tau') \ud \tau' \ud u'-u\tau\\= \int_0^1 I_u(u')I_\tau(u') 
\ud u' -u \tau = \min(u,\tau)-u \tau. \quad
\end{eqnarray}
  Then, the expressions for $\mathbb{E}(B_u^0)^2$ and 
$\mathbb{E}(B_\tau^0)^2$ are readily obtained by setting $u=\tau$ in 
(\ref{eq:A2}). We notice that the above expectations are independent 
of the particular random series representation of the Brownian 
bridge, as they should be.

\section{}
In this appendix, we prove that for the Wiener-Fourier series we have
\begin{equation}
\label{eq:B1}
\lim_{n \to \infty} \frac{\int_0^1 \int_0^1 \gamma_n(u,\tau)^2 
h(u,\tau)\ud u \ud \tau }{\int_0^1 \int_0^1 \gamma_n(u,\tau)^2 \ud u 
\ud \tau} = \int_0^1  h(u,u) \ud u
\end{equation}
for all continuous functions $h(u,\tau)$. This is the statement of 
Eq.~(\ref{eq:13}) in Section~II.B. We start with the equations
\begin{equation}
\label{eq:B2}
\gamma_n(u,\tau)=\frac{2}{\pi^2}\sum_{k=n+1}^\infty \frac{\sin(k\pi 
u)\sin(k\pi \tau)}{k^2}
\end{equation}
and
\[
\int_0^1 \int_0^1 \gamma_n(u,\tau)^2 \ud u \ud 
\tau=\frac{1}{\pi^4}\sum_{k=n+1}^\infty \frac{1}{k^4}.
\]
By letting $n \to \infty$ in the sequence of inequalities
\begin{eqnarray}\nonumber
\label{eq:B2a}
\frac{1}{3}\frac{n^3}{(n+1)^3}=n^3 \int_{n+1}^{\infty}x^{-4} \ud 
x\leq \sum_{k=n+1}^\infty {n^3}/{k^4}\\ \leq n^3 
\int_{n}^{\infty}x^{-4} \ud x =\frac{1}{3},
\end{eqnarray}
one deduces that
\begin{equation}
\label{eq:B3}
\lim_{n\to \infty} n^3 \int_0^1 \int_0^1 \gamma_n(u,\tau)^2 \ud u \ud 
\tau=\frac{1}{3\pi^4}.
\end{equation}
Therefore, Eq.~(\ref{eq:B1}) can be reformulated as
\begin{equation}
\label{eq:B4}
  3\pi^4\lim_{n \to \infty}n^3 {\int_0^1 \int_0^1 \gamma_n(u,\tau)^2 
h(u,\tau)\ud u \ud \tau } = \int_0^1  h(u,u) \ud u
\end{equation}

We prove the relation (\ref{eq:B4}) in two steps. The first step is 
to set $h'(u,\tau)=|h(u,\tau)-h(u,u)|$ and show that
\begin{equation}
\label{eq:B5}
  3\pi^4\lim_{n \to \infty}n^3 {\int_0^1 \int_0^1 \gamma_n(u,\tau)^2 
h'(u,\tau)\ud u \ud \tau } = 0.
\end{equation}
For that purpose, pick an arbitrary $\eta > 0$.
Let us notice that $h'(u,\tau)$ is continuous on the compact set 
$[0,1]\times [0,1]$, thus bounded by a constant $M<\infty$ and 
uniformly continuous. The second property implies that  there is 
$1>\epsilon > 0$  such that $|h'(u',\tau')-h'(u,\tau)|<\eta$ whenever 
$(u'-u)^2+(\tau'-\tau)^2<\epsilon^2$. Now, notice that by 
construction $h'(u,u)=0$ for all $0\leq u\leq 1$. Since any point 
$(u,\tau)$ in the set \[I_\epsilon=\{(u,\tau)\in [0,1]\times[0,1]: 
|u-\tau|<\epsilon\}\] is contained in the ball of radius $\epsilon$ 
centered about the point $(u,u)$, by uniform continuity, it follows 
that
$h'(u,\tau)< \eta$ on $I_\epsilon$.

Let us break the integral in Eq.~(\ref{eq:B5}) in two parts: one over 
the set $I_\epsilon$ and the other on its complementary 
$I_\epsilon^c=[0,1]\times[0,1]-I_\epsilon$.
For the integral over the set $I_\epsilon$ we have
\begin{eqnarray}
\label{eq:B6}&& \nonumber
  3\pi^4\limsup_{n \to \infty}n^3 {\int\! \int_{I_\epsilon} 
\gamma_n(u,\tau)^2 h'(u,\tau)\ud u \ud \tau } \\&& \leq \eta\left[ 
3\pi^4\limsup_{n \to \infty}n^3 {\int\! \int_{I_\epsilon} 
\gamma_n(u,\tau)^2 \ud u \ud \tau }\right] \leq \eta.\quad
\end{eqnarray}
For the integral over the set $I_\epsilon^c$, we have the estimate
\begin{eqnarray}
\label{eq:B7}&& \nonumber
  3\pi^4\limsup_{n \to \infty}n^3 {\int\! \int_{I_\epsilon^c} 
\gamma_n(u,\tau)^2 h'(u,\tau)\ud u \ud \tau } \\&& \leq M\left[ 
3\pi^4\limsup_{n \to \infty}n^3 {\int\! \int_{I_\epsilon^c} 
\gamma_n(u,\tau)^2 \ud u \ud \tau }\right].\quad
\end{eqnarray}
We intend to show that the last limit in Eq.~(\ref{eq:B7}) is zero 
and for this purpose, we need to construct a sharp bound of the 
function $\gamma_n(u,\tau)$ on the set $I_\epsilon^c$. It turns out 
that we first need to study the behavior of the functions
\[
\sum_{k=n+1}^\infty \frac{\cos(k \pi u)}{k^2}
\]
on the interval $(0,2)$.

We have 
\begin{eqnarray}
\label{eq:B9}\nonumber 
\sin(\pi u /2)\sum_{k=n+1}^\infty \frac{\cos(k \pi u)}{k^2}= \frac{1}{2} \sum_{k=n+1}^\infty \frac{\sin[(k+1/2) \pi u]}{k^2}\\-\frac{1}{2} \sum_{k=n+1}^\infty \frac{\sin[(k-1/2) \pi u]}{k^2}=-\frac{1}{2}\frac{\sin[(n+1/2) \pi u]}{(n+1)^2}\qquad \\+\frac{1}{2} \sum_{k=n+1}^\infty\left[\frac{1}{k^2}-\frac{1}{(k+1)^2}\right]{\sin[(k+1/2) \pi u]}.\nonumber \qquad
\end{eqnarray}
From Eq.~(\ref{eq:B9}) we learn that 
\begin{eqnarray}
\label{eq:B10}\nonumber 
\sin(\pi u /2)\left|\sum_{k=n+1}^\infty \frac{\cos(k \pi u)}{k^2}\right|\leq  \frac{1}{2(n+1)^2}\\+\frac{1}{2} \sum_{k=n+1}^\infty\left[\frac{1}{k^2}-\frac{1}{(k+1)^2}\right]=\frac{1}{(n+1)^2}.
\end{eqnarray}
Therefore,
\begin{eqnarray}
\label{eq:B13} 
\left|\sum_{k=n+1}^\infty \frac{\cos(k \pi u)}{k^2}\right|\leq  \frac{1}{(n+1)^2}F(u),
\end{eqnarray}
where the function
\[
F(u)=\frac{1}{\sin(\pi u/2)}
\]
is continuous on the interval $(0,2)$, decreasing on the interval 
$(0,1]$ and increasing on the interval $[1,2)$.

Using the bound given by Eq.~(\ref{eq:B13}) together with the inequality $(a-b)^2 \leq 2(a^2+b^2)$, one computes
\begin{widetext}
\begin{eqnarray}
\label{eq:B14} \nonumber
\gamma_n(u,\tau)^2=\frac{1}{\pi^4}\left\{\sum_{k=n+1}^\infty 
\frac{\cos[k\pi(u-\tau)]-\cos[k\pi(u+\tau)]}{k^2}\right\}^2 \\ \leq 
\frac{2}{\pi^4}\left\{\left[\sum_{k=n+1}^\infty 
\frac{\cos(k\pi|u-\tau|)}{k^2}\right]^2+\left[\sum_{k=n+1}^\infty 
\frac{\cos(k\pi|u+\tau|)}{k^2}\right]^2\right\} \\ \leq
\frac{2}{\pi^4(n+1)^4} \left[F( |u-\tau|)^2+F( |u+\tau|)^2\right].\nonumber
\end{eqnarray}
\end{widetext}
On the set $I_\epsilon^c$, we have $\epsilon\leq|u-\tau|\leq 1$ and 
$\epsilon \leq |u+\tau| \leq 2-\epsilon$. Therefore,  $F( |u-\tau|) 
\leq F(\epsilon)$ and $F( |u+\tau|)\leq F(\epsilon)$. Corroborating 
with Eq.~(\ref{eq:B14}), we obtain
\begin{equation}
\label{eq:B15}
\gamma_n(u,\tau)^2\leq
\frac{4}{\pi^4(n+1)^4}F( \epsilon)^2\quad \text{if}\ (u,\tau)\in I_\epsilon^c.
\end{equation}
Replacing the estimate (\ref{eq:B15}) in Eq.~(\ref{eq:B7}), one 
readily concludes that the limit in Eq.~(\ref{eq:B7}) is 0. Adding 
the limits of Eq.~(\ref{eq:B6}) and Eq.~(\ref{eq:B7}), we conclude 
that
\begin{equation}
\label{eq:B16}
  3\pi^4\limsup_{n \to \infty}n^3 {\int_0^1 \int_0^1 
\gamma_n(u,\tau)^2 h'(u,\tau)\ud u \ud \tau } \leq \eta.
\end{equation}
Since $\eta>0$ is arbitrary, the equality (\ref{eq:B5}) is demonstrated.

In the second step of the proof, we show that Eq.~(\ref{eq:B5}) 
implies the relation (\ref{eq:B1}). From Eq.~(\ref{eq:B5}), one 
deduces that
\begin{eqnarray}
\label{eq:B17}\nonumber
  3\pi^4\lim_{n \to \infty}n^3 {\int_0^1 \int_0^1 \gamma_n(u,\tau)^2 
h(u,\tau)\ud u \ud \tau } \\= 3\pi^4\lim_{n \to \infty}n^3 {\int_0^1 
\int_0^1 \gamma_n(u,\tau)^2 h(u,u)\ud u \ud \tau }.
\end{eqnarray}
By explicitly computing the integral over $\tau$, the last relation becomes
\begin{equation}
\label{eq:B18}
  3\pi^4\lim_{n \to \infty}n^3 \int_0^1  f_n(u) h(u,u)\ud u,
\end{equation}
where
\begin{equation}
\label{eq:B19}
f_n(u)=\frac{2}{\pi^4}\sum_{k=n+1}^\infty \frac{\sin(k\pi 
u)^2}{k^4}=\frac{1}{\pi^4}\sum_{k=n+1}^\infty \frac{1-\cos(2k\pi 
u)}{k^4}.
\end{equation}
By continuity, $h(u,u)$ is integrable and if we set \[B_k=\int_0^1 
{\cos(2k\pi u)} h(u,u)\ud u,\]the Riemann-Lebesgue lemma says that
\[\lim_{n \to \infty} |B_k|=\limsup_{n \to \infty} |B_k|=0.\]
We notice that \begin{eqnarray*}
3\pi^4\lim_{n \to \infty} \frac{n^3}{\pi^4}\left|\sum_{k=n+1}^\infty 
\int_0^1 \frac{\cos(2k\pi u)}{k^4} h(u,u)\ud u \right| \\ \leq 
3\pi^4\lim_{n \to \infty}\left[ 
\frac{n^3}{\pi^4}\left(\sum_{k=n+1}^\infty 
\frac{1}{k^4}\right)\sup_{k\geq n+1} |B_k|\right] \\  \leq \lim_{n\to 
\infty} \sup_{k\geq n+1} |B_k| =0.  \end{eqnarray*}
Therefore, the terms containing $\cos(2k\pi u)$ in the 
Eq.~(\ref{eq:B19}) may be dropped and the quantity~(\ref{eq:B18}) 
becomes
\begin{equation}
\label{eq:B20}
  3\pi^4\lim_{n \to 
\infty}\frac{n^3}{\pi^4}\left(\sum_{k=n+1}\frac{1}{k^4}\right) 
\int_0^1  h(u,u)\ud u  = \int_0^1 h(u,u)\ud u,
\end{equation}
where we used (\ref{eq:B3}) to compute the last limit.
The relation (\ref{eq:B1}) is therefore proved.

\section{}
It is well known that if $A, B >0$, and $\alpha=C/\sqrt{AB}$ such 
that $|\alpha|<1$, then the following Mehler's formula\cite{Sze75} 
holds for all $f$ and $g$ whose squares have finite Gaussian 
transforms
\begin{widetext}
\begin{eqnarray}
\label{eq:C1}
\overline{[fg]}_{ABC}(x_0,y_0)&=&
\int_{\mathbb{R}}\ud x \!\int_{\mathbb{R}} \ud y \frac{1}{2 \pi} 
\frac{1}{\sqrt{AB-C^2}} \exp\left(-\frac{1}{2}\frac{x^2 B + y^2 A -2 
x y C}{AB-C^2}\right)f(x_0+x)g(y_0+y)\nonumber \\ 
&=&\int_{\mathbb{R}}\ud x \!\int_{\mathbb{R}} \ud y \frac{1}{2 \pi} 
\frac{1}{\sqrt{1-\alpha^2}} \exp\left(-\frac{1}{2}\frac{x^2  + y^2 
-2 x y \alpha}{1-\alpha^2}\right)f(x_0+x\sqrt{A})g(y_0+y\sqrt{B}) 
\\&=&
\frac{1}{2\pi} \sum_{k=0}^{\infty} {\alpha^k} \int_{\mathbb{R}}\ud x 
\!\int_{\mathbb{R}} \ud y \,e^{-(x^2+y^2)/2} \,H_k(x) H_k(y) 
f(x_0+x\sqrt{A})g(y_0+y\sqrt{B}).\nonumber
\end{eqnarray}
\end{widetext}
In the above, the functions $H_k(x)$ are the normalized Hermite 
polynomials corresponding to the Gaussian weight
\[\ud \mu(x)=\frac{1}{\sqrt{2\pi}}e^{-x^2/2}.\] They form a complete 
orthonormal basis in the Hilbert space $L^2_{\mu}(\mathbb{R})$, which 
is endowed with the scalar product
\[\langle \psi| \phi\rangle =\int_{\mathbb{R}} \psi(x)\phi(x)\ud 
\mu(x).\] Let us notice that  according to our hypothesis, the 
functions $f(x_0+x\sqrt{A})$ and $g(y_0+x\sqrt{B})$ as functions of 
$x$ are square integrable against $\ud \mu(x)$ and thus they lie in 
the Hilbert space $L^2_{\mu}(\mathbb{R})$.

By repeated integration by parts, the formula (\ref{eq:C1}) is shown to equal
\begin{equation}
\label{eq:C2}
\overline{[fg]}_{ABC}(x_0,y_0)=\sum_{k=0}^\infty \frac{C^k}{k!} 
\overline{f}_A^{(k)}(x_0) \overline{g}_B^{(k)}(y_0),
\end{equation}
where in general $\overline{f}_A^{(k)}(x_0)$ is the $k$-order derivative of
\[\overline{f}_A(x_0)=\int_{\mathbb{R}} \frac{1}{\sqrt{2\pi A}} 
e^{-z^2/(2A)}f(x_0+z)\ud z.\]
Let us notice that the series (\ref{eq:C1}) can be extended to the 
case $\alpha=1$, too. Indeed, the last series in Eq.~(\ref{eq:C1}) 
for the case $\alpha=1$ is nothing else but the Bessel series 
\[\sum_{k=0}^\infty \left\langle H_k|f(x_0+ \cdot 
\sqrt{A})\right\rangle \left\langle H_k|g(y_0+ \cdot 
\sqrt{B})\right\rangle,\]
which is convergent to
\begin{eqnarray*}&&
\left\langle f(x_0+ \cdot \sqrt{A})|g(y_0+ \cdot 
\sqrt{B})\right\rangle \\&& = \int_{\mathbb{R}}f(x_0+ x 
\sqrt{A})g(y_0+x \sqrt{B})\ud \mu(x).
\end{eqnarray*}

Next, we proceed to establish the relation (\ref{eq:17}).
We start with the identity
\begin{eqnarray}
\label{eq:C3}&&
\mathbb{E}_n \left[U_\infty(x,x',\beta;\bar{a})- 
\mathbb{E}_n\,U_\infty(x,x',\beta;\bar{a})\right]^2 \nonumber \\&& 
=\mathbb{E}_n\, U_\infty(x,x',\beta;\bar{a})^2- 
\left[\mathbb{E}_n\,U_\infty(x,x',\beta;\bar{a})\right]^2 .\quad
\end{eqnarray}
Using the notation introduced in Section~II.A, we have
\begin{equation}
\label{eq:C4}
\mathbb{E}_n 
U_\infty(x,x',\beta;\bar{a})=\overline{V}_{u,n}[x_r(u)+\sigma 
S^n_u(\bar{a})].
\end{equation}
Moreover,
\begin{eqnarray}
\label{eq:C5}&&
\mathbb{E}_n U_\infty(x,x',\beta;\bar{a})^2\nonumber =\int_0^1\! \ud 
u \!\int_0^1 \!\ud \tau
\mathbb{E}_n  V[x_r(u)\\&&+\sigma S_u^n(\bar{a})+\sigma 
B_u^n(\bar{a})]V[x_r(\tau)+\sigma S_\tau^n(\bar{a})+\sigma 
B_\tau^n(\bar{a})]\qquad
\end{eqnarray}
and remembering that the variables $B_u^n(\bar{a})$ and 
$B_\tau^n(\bar{a})$ have a joint Gaussian distribution of covariances 
given by Eq.~(\ref{eq:7}), we obtain
\begin{eqnarray*}&&
\mathbb{E}_n U_\infty(x,x',\beta;\bar{a})^2=\int_0^1\! \ud u 
\!\int_0^1 \!\ud \tau
\int_{\mathbb{R}}\ud x \int_{\mathbb{R}} \ud y \frac{1}{2\pi 
\Delta_n(u,\tau)}\nonumber \\&& \times 
\exp\left\{-\frac{1}{2}\frac{x^2\Gamma^2_n(\tau)+y^2\Gamma^2_n(u)-2xyG_n(u,\tau)}{\Delta^2_n(u,\tau)}\right\} 
\\ && \times  V[x_r(u)+\sigma S_u^n(\bar{a})+x]V[x_r(\tau)+\sigma 
S_\tau^n(\bar{a})+y],\qquad
\end{eqnarray*}
where  $G_n(u,\tau)=\sigma^2\gamma_n(u,\tau)$ and
\[\Delta^2_n(u,\tau)=\Gamma^2_n(u)\Gamma^2_n(\tau)-G_n(u,\tau)^2.\]

Using the expansion (\ref{eq:C2}), one may write the above integral 
as the sum of the series
\begin{eqnarray*}
\nonumber
\mathbb{E}_n 
U_\infty(x,x',\beta;\bar{a})^2=\sum_{k=0}^{\infty}\frac{1}{k!}\int_0^1\! 
\ud u \!\int_0^1 \!\ud \tau G_n(u,\tau)^k \\ \times 
\overline{V}_{u,n}^{(k)}[x_r(u)+\sigma 
S_u^n(\bar{a})]\overline{V}_{\tau,n}^{(k)}[x_r(\tau)+\sigma 
S_\tau^n(\bar{a})],
\end{eqnarray*}
where $\overline{V}_{u,n}^{(k)}(x)$ is the $k$ order derivative of 
$\overline{V}_{u,n}(x).$ With the help of Eq.~(\ref{eq:C4}), one 
recognizes the first term of the above series to be $[\mathbb{E}_n 
U_\infty(x,x',\beta;\bar{a})]^2$, so that Eq.~(\ref{eq:C3}) becomes
\begin{eqnarray}
\label{eq:C6}
\mathbb{E}_n \left[U_\infty(x,x',\beta;\bar{a})- \mathbb{E}_n 
U_\infty(x,x',\beta;\bar{a})\right]^2 \nonumber \\ 
=\sum_{k=1}^{\infty}\frac{1}{k!}\int_0^1\! \ud u \!\int_0^1 \!\ud 
\tau G_n(u,\tau)^k \qquad \\ \times 
\overline{V}_{u,n}^{(k)}[x_r(u)+\sigma 
S_u^n(\bar{a})]\overline{V}_{\tau,n}^{(k)}[x_r(\tau)+\sigma 
S_\tau^n(\bar{a})].\nonumber
\end{eqnarray}
As mentioned in Section~II.B, the kernels $\gamma_n(u,\tau)^k$ and 
therefore $G_n(u,\tau)^k$ are positive definite for all integers 
$k\geq 1$ so that the terms of the series (\ref{eq:C6}) are all 
positive i.e.,
  \begin{eqnarray*}
\int_0^1\! \ud u \!\int_0^1 \!\ud \tau G_n(u,\tau)^k 
\overline{V}_{u,n}^{(k)}[x_r(u)+\sigma S_u^n(\bar{a})]\\ 
\overline{V}_{\tau,n}^{(k)}[x_r(\tau)+\sigma S_\tau^n(\bar{a})]\geq 
0\end{eqnarray*} for all $k\geq 1$.
Because of this property, the series (\ref{eq:C6}) will plays the 
fundamental role in establishing the various results in this appendix.
Truncating the series (\ref{eq:C6}) to the first term, we obtain the inequality
\begin{equation}
\label{eq:C7}
\mathbb{E}_n \left[U_\infty(x,x',\beta;\bar{a})- 
\mathbb{E}_n\,U_\infty(x,x',\beta;\bar{a})\right]^2 \geq \mathbb{E}_n 
T'_n(x,x',\beta;\bar{a}),
\end{equation}
where
\begin{eqnarray*}&&T'_n(x,x',\beta;\bar{a})= \int_0^1\! \ud u 
\!\int_0^1 \!\ud \tau G_n(u,\tau) \\&& \times 
\overline{V}_{u,n}^{(1)}[x_r(u)+\sigma S^n_u(\bar{a})] 
\overline{V}_{\tau,n}^{(1)}[x_r(\tau)+\sigma 
S^n_\tau(\bar{a})].\end{eqnarray*}
In a similar fashion, if we truncate the series (\ref{eq:C6}) to the 
first two terms, we obtain
\begin{equation}
\label{eq:C8}
\mathbb{E}_n \left[U_\infty(x,x',\beta;\bar{a})- 
\mathbb{E}_n\,U_\infty(x,x',\beta;\bar{a})\right]^2 \geq \mathbb{E}_n 
Y'_n(x,x',\beta;\bar{a}),
\end{equation}
where
\begin{eqnarray*}Y'_n(x,x',\beta;\bar{a})=T'_n(x,x',\beta;\bar{a})+\frac{1}{2} 
\int_0^1\! \ud u \!\int_0^1 \!\ud \tau G_n(u,\tau)^2 \\ \times 
\overline{V}_{u,n}^{(2)}[x_r(u)+\sigma S^n_u(\bar{a})] 
\overline{V}_{\tau,n}^{(2)}[x_r(\tau)+\sigma 
S^n_\tau(\bar{a})].\end{eqnarray*}

Now, we assume that $V(x) \in \cap_\alpha W^{1,2}_\alpha (\mathbb{R})$.
The inequality $1/k! \leq 1/(k-1)!$ and the positivity of the terms 
of the series (\ref{eq:C6}) imply
\begin{widetext}
\begin{eqnarray}
\label{eq:C9}&&
\mathbb{E}_n \left[U_\infty(x,x',\beta;\bar{a})- 
\mathbb{E}_n\,U_\infty(x,x',\beta;\bar{a})\right]^2 \nonumber   \leq 
\int_0^1\! \ud u \!\int_0^1 \!\ud \tau G_n(u,\tau) 
\bigg\{\sum_{k=0}^{\infty}\frac{1}{k!} G_n(u,\tau)^{k}\\ &&\times 
\overline{V}_{u,n}^{(k+1)}[x_r(u)+\sigma S^n_u(\bar{a})] 
\overline{V}_{\tau,n}^{(k+1)}[x_r(\tau)+\sigma 
S^n_\tau(\bar{a})]\bigg\}= \int_0^1\! \ud u \!\int_0^1 \!\ud \tau 
G_n(u,\tau) \\ &&\nonumber \times \mathbb{E}_n 
\left\{{V}^{(1)}[x_r(u)+\sigma B^0_u(\bar{a})] 
{V}^{(1)}[x_r(\tau)+\sigma B^0_\tau(\bar{a})]\right\}= \mathbb{E}_n 
T_n(x,x',\beta;\bar{a}),
\end{eqnarray}
\end{widetext}
where
\begin{eqnarray*}
T_n(x,x',\beta;\bar{a})=\int_0^1\! \ud u \!\int_0^1 \!\ud \tau 
G_n(u,\tau)   {V}^{(1)}[x_r(u)+\sigma B^0_u(\bar{a})]\\ \times 
{V}^{(1)}[x_r(\tau)+\sigma B^0_\tau(\bar{a})].
\end{eqnarray*}
Similarly, by employing the inequality $1/k! \leq 1/[2(k-1)!]$ if 
$k\geq 2$, one proves the sharper inequality
\begin{equation}
\label{eq:C10}
\mathbb{E}_n \left[U_\infty(x,x',\beta;\bar{a})- 
\mathbb{E}_n\,U_\infty(x,x',\beta;\bar{a})\right]^2 \leq \mathbb{E}_n 
Y''_n(x,x',\beta;\bar{a}),
\end{equation}
where
\[Y''_n(x,x',\beta;\bar{a})=\frac{1}{2}\left[T'_n(x,x',\beta;\bar{a})+T_n(x,x',\beta;\bar{a})\right].\]

If in addition $V(x)\in \cap_\alpha W^{2,2}_\alpha (\mathbb{R})$,
\begin{widetext}
\begin{eqnarray}
\label{eq:C11}\nonumber
\mathbb{E}_n Y''_n(x,x',\beta;\bar{a})-\mathbb{E}_n 
Y'_n(x,x',\beta;\bar{a})=\frac{1}{2}  \int_0^1\! \ud u \!\int_0^1 
\!\ud \tau G_n(u,\tau)^2 \bigg\{\sum_{k=1}^{\infty}\frac{1}{(k+1)!} 
G_n(u,\tau)^{k} \qquad \\ \times 
\overline{V}_{u,n}^{(k+2)}[x_r(u)+\sigma S^n_u(\bar{a})] 
\overline{V}_{\tau,n}^{(k+2)}[x_r(\tau)+\sigma 
S^n_\tau(\bar{a})]\bigg\} \leq \frac{1}{2}  \int_0^1\! \ud u 
\!\int_0^1 \!\ud \tau G_n(u,\tau)^2 
\bigg\{\sum_{k=1}^{\infty}\frac{1}{k!} G_n(u,\tau)^{k}  \\ \times 
\overline{V}_{u,n}^{(k+2)}[x_r(u)+\sigma S^n_u(\bar{a})] 
\overline{V}_{\tau,n}^{(k+2)}[x_r(\tau)+\sigma 
S^n_\tau(\bar{a})]\bigg\}.\nonumber
\end{eqnarray}
\end{widetext}
One recognizes  the last series in Eq.~(\ref{eq:C11}) to be equal to
\begin{eqnarray}
\label{eq:C12}&&\nonumber
\mathbb{E}_n\left\{{V}^{(2)}[x_r(u)+\sigma B_{u}^0(\bar{a})] 
{V}^{(2)}[x_r(\tau)+\sigma B^0_\tau(\bar{a})]\right\} \\&&- 
\overline{V}_{u,n}^{(2)}[x_r(u)+\sigma 
S_{u}^n(\bar{a})]\overline{V}_{\tau,n}^{(2)}[x_r(\tau)+\sigma 
S^n_\tau(\bar{a})]
\end{eqnarray}
and therefore, the inequality ($\ref{eq:40e}$) of Section~III.B is proved.

The multidimensional case is left to the reader and can be obtained 
starting with the following $d$-dimensional analog of 
Eq.~(\ref{eq:C6}). Let $I_k$ be the set of all vectors 
$\bar{k}=(k_1,\ldots,k_d)$ of positive components such that 
$k_1+\cdots k_d=k$ and let $\partial^k_{k_1,\cdots, k_d}f(x)$ denote 
the respective partial derivative of the $d$-dimensional function 
$f(x)$. Then,
\begin{widetext}
\begin{eqnarray}
\label{eq:C13}\nonumber&&
\mathbb{E}_n \left[U_\infty(x,x',\beta;\bar{a})- \mathbb{E}_n 
U_\infty(x,x',\beta;\bar{a})\right]^2 
=\sum_{k=1}^{\infty}\frac{1}{k!}\int_0^1\! \ud u \!\int_0^1 \!\ud 
\tau \gamma_n(u,\tau)^k \qquad \\&& \times \left\{ \sum_{\bar{k}\in 
I_k} \frac{k!}{k_1!\cdots k_d!} \sigma_1^{2k_1}\cdots 
\sigma_d^{2k_d}\left\{\partial^k_{k_1,\ldots, k_d} 
\overline{V}_{u,n}[x_r(u)+\sigma 
S_u^n(\bar{a})]\;\partial^k_{k_1,\ldots, k_d} 
\overline{V}_{\tau,n}[x_r(\tau)+\sigma 
S_\tau^n(\bar{a})]\right\}\right\}.
\end{eqnarray}
\end{widetext}
The equation (\ref{eq:C13}) can be deduced by remembering that the 
Brownian bridges $B_{u,k}^0$ for each dimension are independent 
processes. Then, one successively applies the expansion (\ref{eq:C6}) 
for each dimension to obtain Eq.~(\ref{eq:C13}).

\end{document}